\documentclass[12pt]{article}
\usepackage{fancyhdr}

\usepackage{mathrsfs}
\usepackage[T1]{fontenc}
\usepackage{mathpazo}
\usepackage{setspace}
\usepackage{amsfonts}
\usepackage{amssymb}
\usepackage{amsmath}
\usepackage{epsfig}
\usepackage{latexsym}
\usepackage{color}
\usepackage{graphicx}
\usepackage{nicefrac}
\usepackage[latin1]{inputenc}
\usepackage{slashed}
\usepackage{multirow}
\usepackage{comment}
\usepackage{hyperref}
\usepackage{cite}
\usepackage{soul}

\def\hybrid{\topmargin -20pt    \oddsidemargin 0pt
        \headheight 0pt \headsep 0pt
        \textwidth 6.25in       
        \textheight 9.25in       
        \marginparwidth .875in
        \parskip 5pt plus 1pt   \jot = 1.5ex}

\hybrid

\def\baselinestretch{1.2}

\catcode`\@=11

\def\marginnote#1{}
\newcount\hour
\newcount\minute
\newtoks\amorpm
\hour=\time\divide\hour by60
\minute=\time{\multiply\hour by60 \global\advance\minute by-\hour}
\edef\standardtime{{\ifnum\hour<12 \global\amorpm={am}%
        \else\global\amorpm={pm}\advance\hour by-12 \fi
        \ifnum\hour=0 \hour=12 \fi
        \number\hour:\ifnum\minute<10 0\fi\number\minute\the\amorpm}}
\edef\militarytime{\number\hour:\ifnum\minute<10 0\fi\number\minute}

\def\draftlabel#1{{\@bsphack\if@filesw {\let\thepage\relax
   \xdef\@gtempa{\write\@auxout{\string
      \newlabel{#1}{{\@currentlabel}{\thepage}}}}}\@gtempa
   \if@nobreak \ifvmode\nobreak\fi\fi\fi\@esphack}
        \gdef\@eqnlabel{#1}}
\def\@eqnlabel{}
\def\@vacuum{}
\def\draftmarginnote#1{\marginpar{\raggedright\scriptsize\tt#1}}

\def\draft{\oddsidemargin -.5truein
        \def\@oddfoot{\sl preliminary draft \hfil
        \rm\thepage\hfil\sl\today\quad\militarytime}
        \let\@evenfoot\@oddfoot \overfullrule 3pt
        \let\label=\draftlabel
        \let\marginnote=\draftmarginnote
   \def\@eqnnum{(\theequation)\rlap{\kern\marginparsep\tt\@eqnlabel}%
\global\let\@eqnlabel\@vacuum}  }

\def\preprint{\twocolumn\sloppy\flushbottom\parindent 2em
        \leftmargini 2em\leftmarginv .5em\leftmarginvi .5em
        \oddsidemargin -.5in    \evensidemargin -.5in
        \columnsep .4in \footheight 0pt
        \textwidth 10.in        \topmargin  -.4in
        \headheight 12pt \topskip .4in
        \textheight 6.9in \footskip 0pt
        \def\@oddhead{\thepage\hfil\addtocounter{page}{1}\thepage}
        \let\@evenhead\@oddhead \def\@oddfoot{} \def\@evenfoot{} }

\def\numberbysection{\@addtoreset{equation}{section}
        \def\theequation{\thesection.\arabic{equation}}}

\def\underline#1{\relax\ifmmode\@@underline#1\else
        $\@@underline{\hbox{#1}}$\relax\fi}

\def\titlepage{\@restonecolfalse\if@twocolumn\@restonecoltrue\onecolumn
     \else \newpage \fi \thispagestyle{empty}\c@page\z@
        \def\thefootnote{\fnsymbol{footnote}} }

\def\endtitlepage{\if@restonecol\twocolumn \else \newpage \fi
        \def\thefootnote{\arabic{footnote}}
        \setcounter{footnote}{0}}  

\catcode`@=12
\relax

\def\figcap{\section*{Figure Captions\markboth
        {FIGURECAPTIONS}{FIGURECAPTIONS}}\list
        {Figure \arabic{enumi}:\hfill}{\settowidth\labelwidth{Figure
999:}
        \leftmargin\labelwidth
        \advance\leftmargin\labelsep\usecounter{enumi}}}
 \relax
\def\tablecap{\section*{Table Captions\markboth
        {TABLECAPTIONS}{TABLECAPTIONS}}\list
        {Table \arabic{enumi}:\hfill}{\settowidth\labelwidth{Table
999:}
        \leftmargin\labelwidth
        \advance\leftmargin\labelsep\usecounter{enumi}}}
 \relax
\def\reflist{\section*{References\markboth
        {REFLIST}{REFLIST}}\list
        {[\arabic{enumi}]\hfill}{\settowidth\labelwidth{[999]}
        \leftmargin\labelwidth
        \advance\leftmargin\labelsep\usecounter{enumi}}}
 \relax
\makeatletter
\newcounter{pubctr}
\def\publist{\@ifnextchar[{\@publist}{\@@publist}}
\def\@publist[#1]{\list
        {[\arabic{pubctr}]\hfill}{\settowidth\labelwidth{[999]}
        \leftmargin\labelwidth
        \advance\leftmargin\labelsep
        \@nmbrlisttrue\def\@listctr{pubctr}
        \setcounter{pubctr}{#1}\addtocounter{pubctr}{-1}}}
\def\@@publist{\list
        {[\arabic{pubctr}]\hfill}{\settowidth\labelwidth{[999]}
        \leftmargin\labelwidth
        \advance\leftmargin\labelsep
        \@nmbrlisttrue\def\@listctr{pubctr}}}
 \relax
\makeatother
\newskip\humongous \humongous=0pt plus 1000pt minus 1000pt

\newif\ifdtup

\relax

\def\be{\begin{equation}}
\def\ee{\end{equation}}
\def\ba{\begin{eqnarray}}
\def\ea{\end{eqnarray}}

\def\del{\partial}

\def\r{\rho}
\def\a{\alpha}

\def\d{\delta}

\def\l{\lambda}
\def\L{\Lambda}
\def\s{\sigma}

\def\cN{{\cal N}}

\def\no{\noindent}

\def\qq{\qquad}

\def\IR{\relax{\rm I\kern-.18em R}}

\def \ha {{1\over 2}}

\def \ov {\over}

\def\IR{\relax{\rm I\kern-.18em R}}
\def\IL{\relax{\rm I\kern-.18em L}}

\def\inv{^{\raise.15ex\hbox{${\scriptscriptstyle -}$}\kern-.05em 1}}

\def\tr{{\rm tr}}
\def\Tr{{\rm Tr}}

\begin{document}

\renewcommand{\theequation}{\thesection.\arabic{equation}}
\csname @addtoreset\endcsname{equation}{section}

\newcommand{\beq}{\begin{equation}}
\newcommand{\eeq}[1]{\label{#1}\end{equation}}
\newcommand{\ber}{\begin{eqnarray}}
\newcommand{\eer}[1]{\label{#1}\end{eqnarray}}
\newcommand{\eqn}[1]{(\ref{#1})}
\begin{titlepage}
\begin{center}

\hfill DMUS--MP--14/10\\

\vskip  .7in

{\large \bf The classical Yang--Baxter equation and the associated \\
Yangian symmetry of gauged WZW-type theories}

\vskip 0.4in

{\bf Georgios Itsios},$^{1}$\phantom{x}{\bf {Konstantinos} Sfetsos},$^{2}$\phantom{x}\\
{\bf Konstantinos Siampos}$^{3}$ and {\bf Alessandro Torrielli}$^{4}$
\vskip 0.1in
{\em
\vskip .15in
${}^1$Department of Mathematics, University of Patras,
26110 Patras, Greece\\
{\tt gitsios@upatras.gr}\\
\vskip 0.1in
${}^2$Department of Nuclear and Particle Physics,
Faculty of Physics,\\
University of Athens,
15771 Athens, Greece\\
{\tt ksfetsos@phys.uoa.gr}\\
\vskip 0.1in
${}^3$M\'ecanique et Gravitation, Universit\'e de Mons, 7000 Mons, Belgium\\
{\tt konstantinos.siampos@umons.ac.be}\\
\vskip 0.1in
${}^4$Department of Mathematics, University of Surrey, Guildford GU2 7XH, UK\\
{\tt a.torrielli@surrey.ac.uk}
}

\vskip .4in
\end{center}

\centerline{\bf Abstract}

\no
We construct the Lax-pair, the classical monodromy matrix and the corresponding solution of the Yang--Baxter equation,
{for a two-parameter deformation of the Principal chiral model for a simple group. This deformation includes 
as a one-parameter subset, a class of integrable gauged WZW-type theories
interpolating between the WZW model and the non-Abelian T-dual of the principal chiral model.}
We derive in full detail
the Yangian algebra using two independent methods: by computing the algebra of the non-local charges and alternatively through an expansion of the Maillet
brackets for the monodromy matrix.
As a byproduct, we also provide a detailed general proof of the Serre relations for the Yangian symmetry.

\vskip1.0cm
\hfill {\it Dedicated to the memory of Sotirios Bonanos whose MATHEMATICA software\\ \phantom{xxxxxxxxxxx}has helped numerous researchers}

\newpage

\tableofcontents

\noindent

\vskip .4in

\end{titlepage}
\vfill
\eject

\def\baselinestretch{1.2}
\baselineskip 20 pt
\noindent


\setcounter{equation}{0}
\section{Introduction and motivation}
\label{intro}
\renewcommand{\theequation}{\thesection.\arabic{equation}}

A class of $\s$-models was recently constructed via a gauging procedure involving the
WZW action and the general Principal Chiral model (PCM) action for a group $G$ \cite{Sfetsos:2013wia}.
The end result is the action
\be
\label{sigma.general}
S_{k,\lambda}(g)=S_{\text{WZW},k}(g)+\frac{k}{\pi}\int J^a_+\,(\lambda^{-1}-D^T)^{-1}_{ab}J_-^b\,,
\ee
where $S_{\text{WZW},k}(g)$ is the WZW action at level $k$ of a group element $g\in G$ and
$\lambda$ is a general $\text{dim}(\bf{g})$ square real matrix. In addition, we have employed the standard definitions
\be
J^a_+ ={\rm Tr}(T_a \del_+ g g^{-1}) \ ,\qq J^a_- = \, {\rm Tr}(T_a g^{-1} \del_- g )\ ,\qq D_{ab}={\rm Tr}(T_a g T_b g^{-1})\ ,
\label{jjd}
\ee
with $T_a,\, a=1,2,\dots,\text{dim}(\bf{g})$ being the generators of the Lie algebra {\bf g} satisfying the commutation
rules, normalization and Killing form
\begin{equation*}
[T_a,T_b]=f_{abc}\,T_c\,,\qquad \Tr(T_aT_b)=\delta_{ab}\,,\qquad K^a_b = \delta^a_b\,.
\end{equation*}
The key property of this action arises when $\lambda$ is proportional to the identity, i.e. $\l_{ab}=\l \d_{ab}$,
since then
it becomes integrable. This was shown in \cite{Sfetsos:2013wia} by explicitly demonstrating that the
current components $I_\pm=I_\pm^a T_a$ obey the standard integrability conditions
\be
\label{eomrhozero}
\del_+ I_- +  \del_- I_+ =0\ , \qquad \del_+ I_- - \del_- I_+ +  [I_+,I_-]=0\ .
\ee
The explicit realization in terms of the $\s$-model action variables is
\be
\begin{split}
&H = {1\ov 4 e^2} \int_{-\infty}^{+\infty} \mathrm{d}\s ( I^a_+ I^a_+ + I^a_- I^a_-)\,, \\
&I_+^a=\frac{2\lambda}{1+\lambda}\,(\mathbb{I}-\lambda D)^{-1}_{ab}J^b_+\,,\qquad
I_-^a=-\frac{2\lambda}{1+\lambda}\,(\mathbb{I}-\lambda D^T)^{-1}_{ab}J^b_-\,.
\label{hamml}
\end{split}
\ee
where we have also included the expression for the Hamiltonian corresponding to \eqn{sigma.general}. The general proof was done in \cite{Sfetsos:2013wia} by
explicitly demonstrating that certain integrability algebraic constraints provided in \cite{Balog:1993es,Evans:1994hi} were satisfied.
A simpler way to prove the
integrability of \eqn{sigma.general} has been given more recently in \cite{Hollowood:2014rla} by utilizing the fact that the construction
involves, as mentioned, a gauging procedure reminiscent of the gauged WZW models.

\no
As discussed in detail in \cite{Sfetsos:2013wia}
a motivation for studying this action relates to the global properties of the variables in $\s$-models arising via
non-Abelian T-duality.
The latter generalizes, in a certain sense, Abelian T-duality \cite{Buscher:1987sk}
and was initiated by \cite{Fridling:1983ha,Fradkin:1984ai,delaOssa:1992vc}.
It is easily seen that when the elements $\l_{ab}\to 0$ then \eqn{sigma.general} becomes the WZW $S_{{\rm WZW},k}(g)$.
Also recall \cite{Sfetsos:2013wia} that when $\l$ approaches the identity matrix, $k\to \infty$ and $g\in G$ is
appropriately expanded around the identity group element,
then \eqn{sigma.general} becomes the non-Abelian T-dual for the general PCM.\footnote{For
the isotropic case see the derivation in \cite{Curtright:1994be,Lozano:1995jx} and for the general anisotropic case in \cite{Sfetsos:1996pm}.
Recent developments in non-Abelian T-duality in the presence of RR flux fields initiated with the work in \cite{Sfetsos:2010uq}.
For relations to the AdS/CFT correspondence, a discussion of global issues and
further developments and references see \cite{Itsios:2013wd,Barranco:2013fza,Lozano:2013oma}.}
Hence \eqn{sigma.general} interpolates
between these two extreme cases and a way of thinking to the non-Abelian T-duality of the PCM is as a limiting case of \eqn{sigma.general}.
In the latter action the group element $g\in G$ is parametrized by compact variables. Hence the non-compactness displayed
by the variables in the non-Abelian model is attributed to the zooming-limiting procedure we mentioned.

\no
The perturbation away from the WZW point is driven by the term $\l_{ab} J_+^a J_-^b$ which {  for generic $\l_{ab}$ preserves no isometries} (enhanced to
$G_L\times G_R$ when $\l_{ab}=\l \d_{ab}$).
Based on that, on the matching of global symmetries and on the result of the computation of the renormalization group flow equations
for the matrix $\l$ in \cite{Itsios:2014lca,Sfetsos:2014jfa}, one concludes that \eqn{sigma.general} provides the effective
action for the bosonized anisotropic non-Abelian Thirring model valid to all orders in $\l$ and to leading order in the $1/k$ expansion.
In the same papers the following remarkable symmetry was also noticed
\be
\label{symmetry.sigma}
S_{-k,\l^{-1}}(g^{-1}) = S_{k,\l}(g)\,,
\ee
which in fact mathematically dictates the form of all the aforementioned properties.

In this paper we will further investigate the integrable structure of { a two-parameter deformation of the PCM, which includes
as a one-parameter subset} \eqn{sigma.general} for the prototypical isotropic case $\l_{ab}=\l \d_{ab}$.
In particular, based on the underlying algebraic structure, we will show the existence of a Yangian algebra \cite{Drinfeld:1985rx} 
(for reviews see \cite{Bernard:1992ya,MacKay:2004tc,Torrielli:2010kq}) of classically conserved non-local charges in the
spirit of a similar computation for the (generalized) Gross--Neveu and the isotropic PCM in \cite{MacKay:1992he}. 
{In the isotropic case, the Yangian algebra corresponds to the adjoint action on $g$, i.e. 
$g\mapsto\L_0^{-1} g\L_0\,,\L_0\in G$.}
In addition, we will provide the Lax pair and
we will compute the {Poisson} brackets {of its spatial part which take the Maillet form} \cite{Maillet2,Maillet}. This will provide an array of coefficients which, as required for consistency, solve a classical
modified Yang--Baxter equation. This allows for the derivation of the Maillet brackets of the monodromy matrix \cite{Maillet2,Maillet}.
An expansion of these brackets will provide an alternative derivation of the Yangian algebra.

This work is organized as follows: In section \ref{Lax} we review the derivation of
the Lax pair and the corresponding (classical) monodromy matrix for a general class of two-dimensional systems. In section \ref{Yang} we
compute the Maillet brackets of the spatial part of the Lax pair. Using this, we derive a class of solutions of the modified classical
Yang--Baxter equation and the  Maillet brackets of the monodromy matrix. In section \ref{Yangian} we explore the realization
of the Yangian algebra through the charge algebra and through an expansion of the Maillet brackets of the monodromy matrix. Details of the derivation
are given in Appendices \ref{triple}--\ref{Maillet.expand} respectively. In section \ref{Conclusions}
we conclude with a discussion on possible future directions. Besides Appendices \ref{triple}--\ref{Maillet.expand} we also include
Appendix \ref{Serre.proof} where we revisit the proof of Drinfeld's relations.

\section{Lax pair and the classical monodromy matrix}
\label{Lax}

The purview of this section is to construct the Lax pair and the monodromy matrix of a class of integrable $\s$-models which were constructed in \cite{Sfetsos:2013wia}
and reviewed in section \ref{intro}.
We will provide a rather general discussion by assuming that the
equations of motion and the flat connection identities are given by\footnote{
The world-sheet coordinates $(\s^+,\s^-)$ and $(\tau,\sigma)$ are related by
\begin{equation*}
\s^\pm: = \tau \pm \s \ ,\qq
\del_0 :=\del_\tau= \del_+ +\del_- \ ,\quad \del_1:=\del_\s = \del_+ - \del_- \ ,
\end{equation*}
so that $\star\,\mathrm{d}\s^\pm =\pm \mathrm{d}\s^\pm\, \&\, \star \mathrm{d}\tau = \mathrm{d}\s\ , \star\,\mathrm{d}\sigma = \mathrm{d}\tau$ in Lorentzian signature.
}
\be
\label{eom}
(1+\r) \del_+ I_- + (1-\r) \del_- I_+ =0\,, \qquad
\del_+ I_- - \del_- I_+ +  [I_+,I_-]=0\ ,
\ee
where $I$ is a Lie algebra valued one-form
\be
I=I^aT_a\,,\qquad I^a=I_+^a \mathrm{d}\sigma^++I_-^a\mathrm{d}\sigma^-\ ,
\ee
which for $\rho=0$ describe the integrable (isotropic) $\s$-models reviewed in section \ref{intro}, {whereas for 
$\rho\neq0$ the form of the action is not known.}
We can rewrite \eqref{eom} in a differential-form notation as
\begin{equation}
\label{eom.cal}
{\cal I}:= I - \r \star I\,,\qq \mathrm{d}(\star{\cal I})= 0 \  ,\qq \mathrm{d}{\cal I} + {\cal I}\wedge{\cal I} = 0 \ ,
\end{equation}
{which makes manifest the classical integrability even when $\rho\neq0$.}
Note that, even though the redefinition (\ref{eom.cal})
has made the parameter $\r$ disappear from the integrability conditions, it may very well be present
in the Poisson brackets for {${\cal I}_\pm^a$.}

Using the above and {assuming fields vanish at spatial infinity}, we can construct the first two conserved charges \cite{Luscher:1977rq}
\be
\begin{split}
\label{charges1}
&Q_0 := \, \int_{-\infty}^{+\infty} \mathrm{d}\s\, {\cal I}_0(\s) \,,\\
&\widehat{Q}:= \int_{-\infty}^{+\infty} \mathrm{d}\s\, {\cal I}_1(\s) +\int_{-\infty}^{+\infty} \mathrm{d}\sigma \,
{\cal I}_0(\sigma) \int_{-\infty}^\sigma \mathrm{d}\sigma' \, {\cal I}_0(\sigma'),\
\end{split}
\ee
and another (still conserved and, as we will see, particularly convenient) combination of them
\be
\begin{split}
\label{charges2}
&Q_1:=\widehat Q-\ha\,Q_0^2\,\\
&\phantom{xxx} = \int_{-\infty}^{+\infty} \mathrm{d}\s\, {\cal I}_1(\s)+\ha\int_{-\infty}^{+\infty}
\mathrm{d}\s \int_{-\infty}^\s \mathrm{d}\s'\,[{\cal I}_0(\s),{\cal I}_0(\s')]\,.
\end{split}
\ee
In the above formula, we have rewritten the second term - corresponding to $Q_0^2$ - by splitting the double integral in the two domains ${\s > \s'}$ and ${\s' > \s}$, and changed variables $\s\leftrightarrow\s'$ in one of the pieces.

It is well known that the (infinite number) of conserved charges can be methodically constructed from the {\it Lax pair}
\begin{eqnarray}
\label{flat.Lax}
\partial_0 L_1 - \partial_1 L_0 \, =\, [L_0, L_1]\,\quad\text{or}\quad \mathrm{d}L=L\wedge L\,.
\end{eqnarray}
Using the latter we can show that  the {\it monodromy matrix} (see, for instance, \cite{FTbook})\footnote{
The path ordered exponential reads
\begin{eqnarray*}
P \exp \int_{-\infty}^{+\infty} \mathrm{d}\sigma\,f(\sigma) \, := \, 1 \, + \, \int_{-\infty}^{+\infty}
\mathrm{d}\s\, f(\s) \, + \int_{-\infty}^{+\infty} \mathrm{d}\sigma \int_{-\infty}^\sigma \mathrm{d}\sigma' \, f(\sigma) \, f(\sigma') + \cdots\ .
\end{eqnarray*}
}
\begin{eqnarray}
\label{monodromy}
M (\nu) \,: = \, P \exp \int_{-\infty}^\infty \mathrm{d}\sigma \, L_1(\sigma;\nu)\,
\end{eqnarray}
is conserved for all values of the complex {\it spectral parameter} $\nu$, namely $\partial_0 M (\nu) \, = \, 0.$

In our case the {\it Lax pair} reads
\be
\label{Lax.pair}
L_0 (\sigma;\nu) \, = \, \frac{({\cal I}_1+\tilde\nu\,{\cal I}_0)\tilde\nu}{1-\tilde\nu^2}\,,\quad
L_1 (\sigma;\nu) \, = \frac{({\cal I}_0+\tilde\nu\,{\cal I}_1)\tilde\nu}{1-\tilde\nu^2}\,,\quad
\tilde\nu=\frac{\nu+\rho}{1+\nu\rho}\,,
\ee
or, equivalently, $\displaystyle L_\pm(\sigma;\nu)\,=-\frac{\tilde\nu}{\tilde\nu\mp1}\,{\cal I}_\pm\,.$
By expanding $M$ in powers of $\nu':=\nu+\rho$, we find an infinite set of classically conserved charges:
\begin{equation}
\label{expansion}
M (\nu') \, = 1 \, + \frac{\nu'}{1-\rho^2} \, Q_0 \, +
\frac{{\nu'}^2}{(1-\rho^2)^2} \, \big(\widehat{Q} - \rho Q_0\big)
+ \, {\cal{O}} (\nu'^3)\, .
\end{equation}
One recognizes combinations of the charges \eqref{charges1} in the coefficients of the expansion.

\section{The Maillet brackets and the Yang--Baxter equation}
\label{Yang}

In this section we prove that when the above quantities are supplied by an appropriate algebraic structure, this allows to find explicit solutions
to a modified classical Yang--Baxter equation.
Consequently, the monodromy matrix obeys the associated Maillet brackets.

Following Sklyanin \cite{Evgeny}, we compute the Poisson brackets by first writing $L_1 \, = \, L_1^a \, T_a$ and then\footnote{
\label{tensor.product}
The superscript in parenthesis stands for the notation of tensor products of spaces
\begin{equation*}
\begin{split}
&M^{(1)}=M\otimes\mathbb{I}\,,\qquad M^{(2)}=\mathbb{I}\otimes M\,,\\
&m^{(12)} \, = \, m_{ab} \, T_a \otimes T_b \otimes \mathbb{I}, \quad  m^{(13)} \, = \, m_{ab} \, T_a \otimes \mathbb{I} \otimes T_b, \quad m^{(23)}
\, = \, m_{ab} \, \mathbb{I} \otimes T_a \otimes T_b\ ,
\nonumber
\end{split}
\end{equation*}
for an arbitrary matrix $m=m_{ab}\,T_a\otimes T_b$ in the tensor product algebra.
}
\begin{eqnarray}
\{ L^{(1)}_1 (\sigma_1;\mu), L^{(2)}_1 (\sigma_2;\nu) \} \, = \, \{ L_1^a (\sigma_1;\mu), L_1^b (\sigma_2;\nu) \} \, T_a \otimes T_b\ .
\end{eqnarray}
The Poisson brackets assume the Maillet-type form \cite{Maillet}
\begin{eqnarray}
\label{brac}
\left( \, [r_{-\mu\nu}, L^{(1)}_1 (\sigma_1;\mu)]
+[r_{+\mu\nu},  L^{(2)}_1 (\sigma_1;\nu)]\right)\,\delta_{12} + \delta'_{12} \, \Big(r_{-\mu\nu} - r_{+\mu\nu}\Big)\ ,
\end{eqnarray}
where $r_{\pm\mu\nu}$ (as a shorthand notation for $r_\pm(\mu,\nu)$) are matrices in the basis $T_a\otimes T_b$.
This is guaranteed to give a consistent Poisson structure, provided the Jacobi identities for these brackets are obeyed. This enforces $r_{\pm\mu\nu}$ to satisfy the
modified classical Yang--Baxter relation
\be
\begin{split}
\label{mcYBE}
[r^{(13)}_{+\nu_1\nu_3}, r^{(12)}_{-\nu_1\nu_2}] + [r^{(23)}_{+\nu_2\nu_3}, r^{(12)}_{+\nu_1\nu_2}] +
[r^{(23)}_{+\nu_2\nu_3}, r^{(13)}_{+\nu_1\nu_3}] =0\ .
\end{split}
\ee
The non-vanishing coefficient of the $\d'$ term in \eqn{brac} is responsible for the above modification of the 
classical Yang--Baxter relation.
Using (\ref{brac}), one can derive the Poisson brackets for the (classical) monodromy matrix  \cite{Maillet2}
\be
\begin{split}
\label{Mbrac}
&\{M^{(1)}(\mu), M^{(2)}(\nu)\}\, = \, [r_{\mu\nu}, M(\mu) \otimes M(\nu)]
- M^{(2)}(\nu) \, s_{\mu\nu} \, M^{(1)}(\mu) +\\
&\qquad\qquad\qquad\qquad\qquad M^{(1)}(\mu)\, s_{\mu\nu} M^{(2)}(\nu)\,,
\end{split}
\ee
{
which is consistent with the Jacobi identity,
if we define the equal-point limits of the Poisson brackets through a generalized symmetric 
limit procedure \cite{Maillet2}.
}

Returning to the case at hand, it was pointed out in \cite{Rajeev:1988hq} that the Poisson structure of the isotropic PCM admits a one-parameter family of deformations (with parameter denoted by $x$).
Subsequently, in \cite{Balog:1993es} this deformation was further extended by introducing a second parameter $\r$.
In our conventions such two-parameter algebra reads\footnote{
Alternatively, in light-cone coordinates the algebra reads
\begin{equation*}
\begin{split}
&\{{\cal I}_\pm^a,{\cal I}_\pm^b\}=e^2f_{abc}\left[(1\mp\r)^2{\cal I}_\mp^c-((1\mp\r)^2+2x(1-\r^2)){\cal I}_\pm^c\right]\d_{\s\s'}
\pm2e^2(1\mp\r)^2\d_{ab}\d'_{\s\s'}\ ,
\\
&\{{\cal I}_\pm^a,{\cal I}_\mp^b\}=-e^2f_{abc}\left[(1-\r)^2{\cal I}_-^c+(1+\r)^2{\cal I}_+^c\right]\d_{\s\s'}\ .
\end{split}
\end{equation*}
}
\ba
\label{tilde.iiialfbasis}
&& \{ {\cal I}^a_0 ,  {\cal I}^b_0\} = -2e^2f_{abc}\left(1+\r^2+(1-\r^2)x\right) {\cal I}_0^c\d_{\s\s'}-8e^2\r\,\d_{ab} \d'_{\s\s'}\,,
\nonumber\\
&&  \{  {\cal I}^a_1 ,  {\cal I}^b_1\} = 2e^2f_{abc}\left(4\r\, {\cal I}_1^c+(1+\r^2+x(\r^2-1)) {\cal I}_0^c\right)\d_{\s\s'}-8e^2\r\,\d_{ab} \d'_{\s\s'}\,,\\
&& \{  {\cal I}^a_0 , {\cal I}^b_1\}= \{  {\cal I}^a_1 , {\cal I}^b_0\}  = -2e^2f_{abc}\left(1+\r^2+(1-\r^2)x\right) {\cal I}_1^c\d_{\s\s'}+4e^2(1+\r^2)\d_{ab} \d'_{\s\s'}\,.\nonumber
\ea
When $\r=0$, the action \eqn{sigma.general} provides a realization of this algebra for a general group $G$ \cite{Sfetsos:2013wia}
with $\displaystyle x={\l^2+1\ov 2\l}$
(for the $SU(2)$ case this realization was found by a brute force computation in \cite{Balog:1993es}).
There is no known action realizing the above algebra for $\r\neq 0$.\footnote{\label{foo6} The algebra for the PCM (pseudo-PCM) corresponds to choosing
the parameters $x=1$ ($x=-1$) and $\r=0$. The value $x=1$ corresponds to taking the parameter $\l$ in this paper to unity. The fact that then the action
\eqn{sigma.general} does not become that for a PCM but rather for its non-Abelian T-dual is consistent since non-Abelian T-duality can be
cast as a canonical transformation in phase space \cite{Curtright:1994be,Lozano:1995jx,Sfetsos:1996pm}.
We also note that for $\rho=0$ and $ e\to 0, x\to \infty$, we obtain under an
appropriate rescaling of the currents the
algebra for the (generalized) Gross--Neveu model when $e^2\, x$ is finite, and for the conformal case when $e\,x$ is finite.}
Plugging \eqref{Lax.pair} into \eqref{brac} and using the algebra (\ref{tilde.iiialfbasis}), we find that {the matrix} $r_{\pm\mu\nu}$ read
\begin{equation*}
\begin{split}
&r_{\pm\mu\nu}\mapsto r_{\pm\mu\nu}\,\Pi\,,\qquad \Pi:=\sum_a T^a\otimes T^a\,,\\
&r_{+\mu\nu}=2e^2\frac{(1+\mu^2+x(1-\mu^2))(\mu+\rho)(\nu+\rho)}{(\nu-\mu)(1-\mu^2)}\,,\\
&r_{-\mu\nu}=2e^2\frac{(1+\nu^2+x(1-\nu^2))(\mu+\rho)(\nu+\rho)}{(\nu-\mu)(1-\nu^2)}=-r_{+\nu\mu}\,,
\end{split}
\end{equation*}
{
and henceforth $r_{\pm\mu\nu}$ denotes the scalar.}
As for \eqref{mcYBE}, it reduces to the single algebraic condition
\be
r_{+\nu_2\nu_3}r_{+\nu_1\nu_2}=r_{+\nu_1\nu_3}r_{-\nu_1\nu_2}+r_{+\nu_2\nu_3}r_{+\nu_1\nu_3}\,,
\ee
extracted from the coefficient in front of the combination $f_{abc}\, T^a\otimes T^b\otimes T^c$.
For completeness, we provide the values of $r_{\mu\nu}$ and $s_{\mu\nu}$, obtained by rewriting $r_{\pm\mu\nu}=r_{\mu\nu}\pm s_{\mu\nu}$:
\begin{equation}
\begin{split}
&r_{\mu\nu}=-2e^2\frac{(1-\mu^2\nu^2+x(1-\mu^2)(1-\nu^2))(\mu+\rho)(\nu+\rho)}{(\mu-\nu)(1-\mu^2)(1-\nu^2)}\,,\\
&s_{\mu\nu}=-2e^2\frac{(\mu+\nu)(\mu+\rho)(\nu+\rho)}{(1-\mu^2)(1-\nu^2)}\ ,
\end{split}
\end{equation}
which are generically non-vanishing.

\section{The realization of the Yangian algebra}
\label{Yangian}

The scope of this section is to explicitly realize the Yangian algebra by
performing the mutual Poisson commutators between $Q_0^a$ and $Q_1^a$ and, alternatively, via an expansion of the Maillet brackets for the conserved monodromy matrix $M$.

The Yangian algebra $Y_C\left(\bold{g}\right)$ is an associative Hopf algebra
generated by the elements $J_a$ and $Q_a$ obeying \cite{Drinfeld:1985rx}
\be
\label{Lie}
[J_a,J_b]= F_{abc} J_c\ ,\qq [J_a,Q_b]=[Q_a,J_b]= F_{abc} Q_c\,.
\ee
In addition, the request that the co-product map (which we call $f$ to avoid conflicts of notations) on $J_a$ and $Q_a$, namely
\begin{equation}
\label{coproduct}
f(J_a)=J_a\otimes\mathbb{I}+\mathbb{I}\otimes J_a\,,\quad f(Q_a)=Q_a\otimes\mathbb{I}+\mathbb{I}\otimes Q_a
+\frac{\alpha}{2}\,F_{abc}\,J_b\otimes J_c\,,\quad \alpha\in\mathbb{C}\,,
\end{equation}
acts as a homomorphism\footnote{\label{homo}A homomorphism is a structure-preserving map between two algebraic structures (such as groups)\begin{equation*}
f: A\mapsto B\quad\text{with}\quad f(a_1+a_2)=f(a_1)+f(a_2)\,,\quad f(a_1a_2)=f(a_1)f(a_2)\,,\quad \forall a_1,a_2\in A\,.
\end{equation*}
},
implies the Serre relations - see Appendices \ref{Serre.I} and \ref{Serre.II} for details.
The first Serre relation reads
\begin{equation}
\begin{split}
\label{Drinfeld1}
&[Q_a,[Q_b,J_c]]-[J_a,[Q_b,Q_c]]={\a^2\ov 24} a_{abcdef}  J_{(d}J_eJ_{f)}\,,\\
&a_{abcdef}=F_{adk} F_{bel} F_{cfm} F_{klm}\, ,
\end{split}
\end{equation}
where $J_{(a}J_bJ_{c)}$ denotes the sum of all permutations of $J_aJ_bJ_c$,\footnote{
\label{permute.rule}
This sum explicitly expands as
\begin{equation*}
\label{permute}
J_{(a}J_bJ_{c)}=J_aJ_bJ_c+J_cJ_aJ_b+J_bJ_cJ_a+J_aJ_cJ_b+J_bJ_aJ_c+J_cJ_bJ_a\,.
\end{equation*}} which for (classical)  commuting quantities simplifies to $6\,J_aJ_bJ_c$.
The first Serre relation is trivially satisfied for the $su(2)$ case, as it turns out that $a_{abcdef}=\varepsilon_{abe}\varepsilon_{cfd}-\varepsilon_{dbe}\varepsilon_{cfa}.$
Using the Jacobi identity on the second term of the l.h.s. of \eqn{Drinfeld1}, and using the second of the relations \eqn{Lie}, we easily find that \eqref{Drinfeld1}
can be written as
\be
 F_{dab}[Q_{c},Q_d]+F_{dca}[Q_{b},Q_d] +F_{dbc}[Q_{a},Q_d]  ={\a^2\ov 24} a_{abcdef} J_{(d}J_eJ_{f)}\ ,
\label{Serre}
\ee
a form which is particularly convenient for our purposes.
In addition, the second Serre relation reads
\be
\label{Serre2}
F_{kcd}[[Q_a,Q_b],Q_k]]+F_{kab}[[Q_c,Q_d],Q_k]={\a^2\ov 24} \left(a_{abkgef} F_{kcd} + a_{cdkgef} F_{kab}\right)J_{(g}J_eQ_{f)}\ .
\ee
The first Serre implies the second one (for details see Appendix \ref{Serre.II}), except for the $su(2)$ case where it reads
\be
\begin{split}
\label{Serre3}
&[[Q_a,Q_b],[J_c,Q_d]] + [[Q_c,Q_d],[J_a,Q_b]]=\\
&{\a^2\ov 24}\left(\varepsilon_{eab}(\d_{fc}\d_{gd}-\d_{fd}\d_{cg})+\varepsilon_{ecd}(\d_{fa}\d_{gb}-\d_{fb}\d_{ag})\right)J_{(d}J_eQ_{f)}\ .
\end{split}
\ee
Hence, for the $su(2)$ case only this relation is non-trivial.

\subsection{Yangian algebra through the algebra of charges}

Next we work out the algebra of the classical charges $Q_0$ and $Q_1$ defined in \eqn{charges1} and \eqn{charges2}.
Their components read\footnote{We use the definition
\begin{equation*}
\varepsilon_{{12}}=\varepsilon(\s_1-\s_2)=\left\{
    \begin{array}{ll}
        1  & \mbox{if } \s_1 > \s_2 \\
        -1 & \mbox{if } \s_1 < \s_2
    \end{array}
\right.,\quad \text{and}\quad \varepsilon'(x)=2\delta(x)\,.
\end{equation*}
}
\be
\begin{split}
\label{charges}
& Q_0^a = \int_{-\infty}^{+\infty} \mathrm{d}\s\ {\cal I}_0^a(\s)\ ,\\
&
Q_1^a = \int_{-\infty}^{+\infty} \mathrm{d}\s\ {\cal I}_1^a(\s) + \frac14 f_{abc} \int_{-\infty}^{+\infty}  \int_{-\infty}^{+\infty}
\mathrm{d}^2\s_{12}\,\varepsilon_{{12}}\, {\cal I}_0^b (\s_1){\cal I}_0^c(\s_2)\,.
\end{split}
\ee
A comment is in order regarding the form of the charges. This should be in agreement with the co-product \eqref{coproduct} and realized as
half-positive and half-negative axis splitting (see \cite{MacKay:1992he} for details), upon the identifications
\be
\label{map}
J_a\mapsto Q_0^a\,,\qquad  Q_a\mapsto Q_1^a\,,\qquad F_{abc}\mapsto\frac1\alpha\,f_{abc}\,.
\ee
Using \eqref{tilde.iiialfbasis} we compute the Poisson brackets for the zeroth level charges\footnote{
To avoid ambiguities arising from the non-utralocal terms, like 
\begin{equation*}
\int\mathrm{d}\sigma_1 \mathrm{d}\sigma_2 \del_1\delta_{12}\neq \int\mathrm{d}\sigma_2 \mathrm{d}\sigma_1 \del_1\delta_{12}\,
\end{equation*}
we follow \cite{Luscher:1977rq} and we define the Poisson bracket
\begin{equation*}
\{Q_{0,1}^a,Q_{0,1}^b\}=\lim_{L_2\to\infty}\lim_{L_1\to\infty}\,\{Q_{0,1}^{a,L_1},Q_{0,1}^{b,L_2}\}\,,
\end{equation*}
where $Q^{a,L}_{0,1}$ are volume cutoff charges (the same as $Q^a_{0,1}$, with range from $-L$ to $L$).
}
\be
\label{Q0Q0}
\{ Q_0^a, Q_0^b\}=  -2e^2 \left(1+\rho^2 + x(1- \rho^2)\right)\,f_{abc} Q_0^c\,.
\ee
Using the Jacobi identity we find that
\ba
\label{Q0Q1}
&& \{Q_0^a,Q_1^b\} ={2\Delta}\,f_{abc} \int_{-\infty}^{+\infty} \mathrm{d}\s\ {\cal I}_1^c(\s)
\nonumber\\
&&+{\Delta} ( f_{bef} f_{ae d} +  f_{bde} f_{aef})
 \int^{+\infty}_{-\infty}\mathrm{d}\s_1\ {\cal I}_0^d(\s_1) \int_{-\infty}^{\s_1} \mathrm{d}\s_2\ {\cal I}_0^f(\s_2)\nonumber
 \\
&& \Longrightarrow \{Q_0^a,Q_1^b\} = -2e^2 \left(1+\rho^2 + x(1- \rho^2)\right)\,f_{abc} Q_1^c\ , .
\ea
For notational convenience we define
\begin{eqnarray}
\label{deltadef}
\Delta \, := \, - e^2 \, \left(1+\rho^2 + x(1- \rho^2)\right).
\end{eqnarray}
Finally, we can compute $\{Q_1^a,Q_1^b\}$ as follows:
\ba
\label{Q1Q1partial}
&&Q_1^a:= x^a+y^a\,,\qquad \{Q_1^a,Q_1^b\}=\{x^a,x^b\}+\{x^a,y^b\}-\{x^b,y^a\}+\{y^a,y^b\}\ ,
\nonumber\\
&&\{x^a,x^b\}+\{x^a,y^b\}-\{x^b,y^a\}=f_{abc}\,Q_2^{(1)c}\ ,
\\
&&Q_2^{(1)a}=2e^2(1+\r^2+x(\r^2-1))\,Q_0^a+8e^2\rho\int\mathrm{d}\sigma\,{\cal I}_1^a +
\nonumber\\
&&\qquad \qquad \qquad \qquad \qquad \qquad \qquad \qquad \qquad + \, \Delta\,f_{abc}\int \mathrm{d}^2\s_{12}\,\varepsilon_{{12}}{\cal I}_0^b(\s_1){\cal I}_1^c(\s_2)\ ,
\nonumber
\ea
where $x^a$ and $y^a$ correspond to the first and second term in the expression of $Q_1^a$, respectively.
Furthermore, we can show that
\be
\begin{split}
\label{Poisson.y}
&\{y^a,y^b\}={2}\Delta\,f_{acd}\,f_{bre}f_{dr\ell}\times \frac14\,I_{ce\ell}\,,\\
&I_{ce\ell}=\int \mathrm{d}^3\s_{123}\,\varepsilon_{{13}}\varepsilon_{{32}}{\cal I}_0^c(\s_1){\cal I}_0^e(\s_2){\cal I}_0^\ell(\s_3)=I_{ec\ell}\,,
\end{split}
\ee
which can be further simplified with the use of \eqn{B.9}, proven in Appendix \ref{triple} and reported here
for convenience:
\begin{equation*}
f_{acd}f_{bre}f_{dr\ell}I_{ce\ell}=-\frac13\,f_{acd}f_{bre}f_{dr\ell}\,Q_0^{c}Q_0^{e}Q_0^{\ell}-\frac13\,f_{abr}f_{rcd}f_{de\ell}I_{ce\ell}\,.
\end{equation*}
Putting all together, the Poisson brackets of two $Q_1^a$'s read
\be
\begin{split}
\label{Q1s}
&\{Q_1^a,Q_1^b\}=f_{abc} \left(Q_2^{(1)c}+Q_2^{(2)c}\right)-2\Delta\times f_{acd}\,f_{bre}f_{dr\ell}\times\frac{1}{12} Q_0^cQ_0^eQ_0^\ell\, ,\\
&\text{where}\quad Q_2^{(2)a}=-\frac{\Delta}{6}\,f_{acd}f_{de\ell}\,I_{ce\ell}\,.
\end{split}
\ee
By use of the Jacobi identity, we find the first Serre relation \eqn{Serre}:
\be
\label{SerreI}
\frac12\,f_{d[ab}\{Q_1^{c]},Q_1^d\}=2e^2\left(1+\r^2+(1-\r^2)x\right)\times \frac{1}{24}\,f_{aip}f_{bjq}f_{ckr}f_{ijk}\, Q_0^{(p}Q_0^qQ_0^{r)}\,,
\ee
where we have used the identity \eqref{A.1} proven in Appendix \ref{ident}
\begin{equation}
\label{fiden}
\frac12\,f_{d[ab}f_{c]pm}f_{dnq}f_{mnr}\,Q_0^pQ_0^qQ_0^r=3 f_{aip}f_{bjq}f_{ckr}f_{ijk}\, Q_0^pQ_0^qQ_0^r\,.
\end{equation}
In total, the charges \eqn{charges} form a classical Yangian algebra in the sense of Poisson brackets, namely eqs. \eqref{Q0Q0}, \eqref{Q0Q1}
and \eqref{SerreI}, under the correspondence \eqref{map} with
\be
\label{map1}
\alpha=\frac{1}{2\Delta}=-\frac{1}{2e^2\left(1+\r^2+(1-\r^2)x\right)}\,.
\ee
The above is a generalization of a proof originally given in \cite{MacKay:1992he} for the isotropic PCM and for the
(generalized) Gross--Neveu model, whose corresponding algebras for the $I^a_\pm$'s are particular cases of \eqn{tilde.iiialfbasis}
(see footnote \ref{foo6}). 

The appearance of the classical Yangian algebra was guaranteed in the first place by the existence
of the Yang--Baxter equation \eqref{mcYBE} and the realization of the co-product \eqref{coproduct} by \eqref{charges} and \eqref{map}.
The only additional step which needed to be made was to compute the value of $\alpha$ through the Poisson brackets of the level-zero charges \eqref{Q0Q0}.

\subsubsection{The $su(2)$ case}
\label{Serre.su2}
For the  $su(2)$ case, the first Serre relation is trivially satisfied, therefore we only have to study the second one \eqref{Serre3}.
Using \eqn{Q0Q1} we can rewrite the l.h.s. of \eqn{Serre3} (once again understood in its classical version of Poisson brackets) as
\be
2\Delta\,\left(\varepsilon_{cde}\{\{Q_1^a,Q_1^b\},Q_1^e\}+\varepsilon_{abe}\{\{Q_1^c,Q_1^d\},Q_1^e\}\right)\,.
\ee
Next we note that \eqref{Q1s} trivializes to
\be
\begin{split}
\label{Q1ssu2}
&\{Q_1^a,Q_1^b\}=\varepsilon_{abc}\, {\cal Q}^c\,,\\
&{\cal Q}^a=Q_2^{(1)a}+\frac{\Delta}{4}\,(Q_0^aQ_0^dQ_0^d+I_{dda})\,,
\end{split}
\ee
where $Q_2^{(1)a}$ is given in \eqref{Q1Q1partial} with $f_{abc}$ replaced by $\varepsilon_{abc}$.
Using these specialized expressions we find
\be
\begin{split}
&\{\{Q_1^a,Q_1^b\},\{Q_0^c,Q_1^d\}\} +\{\{Q_1^c,Q_1^d\},\{Q_0^a,Q_1^b\}\} =\\
&2e^6\left(1+\r^2+(1-\r^2)x\right)^3Q_0^e\left(\varepsilon_{abe}\,Q_0^{[c}Q_1^{d]}+\varepsilon_{cde}\,Q_0^{[a}Q_1^{b]}\right)\,,
\end{split}
\ee
which is in agreement with \eqref{Serre3}, \eqref{map} and \eqref{map1}.

\subsection{Yangian algebra through the Maillet brackets of the monodromy matrix}
\label{Maillet}
An alternative derivation of the Yangian algebra is obtained through
the Maillet brackets \eqref{Mbrac} and the expansion of the monodromy matrix $M$ \eqref{expansion}.

Rewriting \eqref{expansion} in terms of $Q_{0,1}$ we find that
\be
\label{expansionn}
M (\nu') \, = 1 \, + \frac{\nu'}{1-\rho^2} \, Q_0 \, +
\frac{{\nu'}^2}{(1-\rho^2)^2} \, \big(Q_1 - \rho Q_0+\ha Q_0^2\big)
+ \, {\cal{O}} (\nu'^3)\,.
\ee
Plugging \eqref{expansionn} into \eqref{Mbrac}, expanding first in $\nu'$ and only afterwards in $\mu'$,
where $\mu':=\mu+\rho$, and keeping all the terms up to the order ${\cal{O}}(\nu'^2 \mu')$, produces,
after a good deal of algebra (and writing $Q = Q^a \, T_a$), \eqref{Q0Q0} and \eqref{Q0Q1}, respectively
\begin{equation}
\label{Maillet.lowest}
\begin{split}
&\{{Q}_0^a, {Q}_0^b\} \, = -2e^2\left(1+\r^2+(1-\r^2)x\right)\, f_{abc} Q_0^c, \\
& \{Q_0^a,Q_1^b\} \, = -2e^2\left(1+\r^2+(1-\r^2)x\right)\, f_{abc} Q_1^c.
\end{split}
\end{equation}
Next we consider the expansion of the Maillet brackets (\ref{Mbrac}) up to ${\cal{O}}(\nu'^2 \mu'^2)$.
We will see that, in order to study this term, it is necessary to expand the monodromy matrix up
to the order ${\cal{O}}(\nu'^3)$. By manipulating the ${\cal{O}}(\nu'^2 \mu'^2)$ term in the brackets, after a rather 
tedious computation, we obtain the first Serre relation in \eqref{SerreI}. The related technical details are presented in Appendix \ref{Maillet.expand}.

\section{Conclusions and outlook}
\label{Conclusions}

The purview of the present paper is the construction of the Lax pair $L_{0,1}$ for isotropic coupling matrices $\lambda$ of the action
\eqref{sigma.general} and the corresponding symmetry algebra \eqref{tilde.iiialfbasis} with $\rho=0$. Using its spatial part $L_1$
we built the conserved classical monodromy matrix,
derived the corresponding Poisson ({\it Maillet-type} \cite{Maillet,Maillet2}) brackets and
the emerging {modified} Yang--Baxter equation as the Jacobi identity on these Poisson brackets. Employing the classical monodromy matrix
we constructed the first two conserved charges and obtained their Yangian algebra, both through the charge algebra and also
from an expansion of the Poisson brackets for the monodromy matrix.
In addition, the renormalizability of this action at one-loop in the $1/k$ expansion \cite{Itsios:2014lca} ensures that
the above construction remains applicable at this order.

It would be interesting to study generalizations of the construction we have provided for vanishing $\rho$ and anisotropic coupling matrices $\lambda$,
whose action was given in \eqref{sigma.general}.
These $\sigma$-models generically interpolate from the WZW to the non-Abelian T-dual of the anisotropic PCM, and so a good place to start
this study are cases which possess an integrable anisotropic PCM endpoint, like the $su(2)$ case \cite{Cherednik:1981df,Hlavaty,Mohammedi:2008vd}.
{  Also note that the Yangian symmetries are 
preserved for the deformed WZW model on squashed spheres \cite{Kawaguchi:2011mz,Kawaguchi:2013gma}.}

It is possible to replace the
WZW term in \eqref{sigma.general} by a coset CFT with
action realization in terms of a gauged WZW model.
In these case the end point of the deformation corresponds to the non-Abelian T-dual of
PCM for coset instead of group spaces \cite{Sfetsos:2013wia}.
In that respect,
and for symmetric coset spaces, the deformation has been
convincingly argued to correspond to a quantum deformation of the
bosonic sector of the string theory, when the deformation parameter
is a root of unity \cite{Hollowood:2014rla}. When instead it is
real, achieved by analytic continuation, the models are those of
\cite{Delduc:2013fga,Delduc:2013qra,Delduc:2014kha}, based on
the construction of \cite{Klimcik:2002zj,Klimcik:2008eq} and realized as
$\sigma$-models in \cite{Arutyunov:2013ega}. 
We believe that our treatment is generalizable to these cases as well.

\section*{Acknowledgements}

We would like to thank Nicolas Boulanger, { Ben Hoare, Marc Magro and Vidas Regelskis} for useful correspondence.
The research of G. Itsios has been co-financed by the ESF (2007-2013) and Greek
national funds through the Operational Program ''Education and
Lifelong Learning" of the NSRF - Research Funding Program:
``Heracleitus II. Investing in knowledge in society through the
European Social Fund". The research of K.\,Sfetsos is implemented
under the \textsl{ARISTEIA} action (D.654 of GGET) of the \textsl{operational
programme education and lifelong learning} and is co-funded by the
European Social Fund (ESF) and National Resources (2007-2013). The work of K.
Siampos has been supported by  \textsl{Actions de recherche
concert\'ees (ARC)} de la \textsl{Direction g\'en\'erale de
l'Enseignement non obligatoire et de la Recherche scientifique
- Direction de la Recherche scientifique - Communaut\'e fran\c{c}aise de
Belgique} (AUWB-2010-10/15-UMONS-1), and by IISN-Belgium (convention 4.4511.06). A.Torrielli thanks EPSRC for funding under the First Grant
project EP/K014412/1 ''Exotic quantum groups, Lie superalgebras and integrable systems". K. Sfetsos and K. Siampos would like to thank the University of Patras
for hospitality, where part of this work was developed. 
A. Torrielli acknowledges useful conversations with the participants of the ESF and STFC supported workshop 
``Permutations and Gauge String duality (STFC-4070083442)'' (Queen Mary U. of London, July 2014).

\appendix

\section{The Serre relations}
\label{Serre.proof}

The scope of this Appendix is to provide an explicit proof of the Serre relations for pedagogical reasons.

\subsection{The first Serre relation}

\label{Serre.I}

The proof goes along the lines suggested in \cite{MacKay:2004tc}. Let us define the co-products of $J_a$ and $Q_a$ as in \eqref{coproduct},
and the quantity
\begin{equation}
\label{D.2}
Z_{ab}:=f([Q_a,Q_b])-[Q_a,Q_b]\otimes\mathbb{I}-\mathbb{I}\otimes[Q_a,Q_b]\,,
\end{equation}
on which $f$ acts as a homomorphism (see footnote \ref{homo}).
Next we introduce
\begin{equation}
\label{D.3}
u_{ab}:=F_{cda}v_{cdb}-F_{cdb}v_{cda}\,,\qquad F_{abc}u_{ab}=0\,,
\end{equation}
where $v_{abc}$ is totally antisymmetric. Contracting \eqn{D.2} with $u_{ab}$, using \eqn{Lie} and
the Jacobi identity on the term proportional to $\alpha$, we find that
\begin{equation}
\label{D.4}
u_{ab}Z_{ab}=\frac{\alpha}{2}\,u_{ab}\,F_{abe}F_{cde}\left(Q_c\otimes J_d-J_c\otimes Q_d\right)+
\frac{\alpha^2}{4}\,u_{ab}F_{acd}F_{bmn}F_{cmr}\left(J_r\otimes J_dJ_n+J_dJ_n\otimes J_r\right)
\end{equation}
with the first term vanishing due to \eqn{D.3}. Substituting the value of $u_{ab}$ we find
\begin{equation}
\label{D.5}
\begin{split}
&\frac{\alpha^2}{4}\,\left(F_{ija}v_{ijb}-F_{ijb}v_{ija}\right)\,F_{acd}F_{bmn}F_{cmr}\left(J_r\otimes J_dJ_n+J_dJ_n\otimes J_r\right)=\\
&\left(A-B\right)\times\frac{\alpha^2}{4}\left(J_r\otimes J_dJ_n+J_dJ_n\otimes J_r\right)\,,
\end{split}
\end{equation}
where
\begin{equation}
\label{D.6}
\begin{split}
&A=v_{ijb}F_{ija}F_{acd}F_{bmn}F_{cmr}=2v_{ijb}\left(F_{ajd}F_{bmn}F_{cmi}F_{car}+F_{ajd}F_{bmn}F_{cam}F_{cir}\right),\\
&B=v_{ija}F_{ijb}F_{acd}F_{bmn}F_{cmr}=2v_{ijb}\left(F_{ajn}F_{bcd}F_{mci}F_{amr}+F_{ajn}F_{bcd}F_{mac}F_{imr}\right),
\end{split}
\end{equation}
and we employed the Jacobi identity twice for each term.
We then rewrite $A-B$ as
\begin{equation}
\label{D.7}
\begin{split}
&A-B=C+D\,,\\
&C=2v_{ijb}\left(F_{ajd}F_{bmn}F_{cmi}F_{car}-F_{ajn}F_{bcd}F_{mci}F_{amr}\right),\\
&D=2v_{ijb}\left(F_{ajd}F_{bmn}F_{cam}F_{cir}-F_{ajn}F_{bcd}F_{mac}F_{imr}\right).
\end{split}
\end{equation}
Applying the Jacobi identity on $C$ and relabelling the indices of $D$, we find
\begin{equation}
\label{D.8}
D=4v_{ijb}F_{ajd}F_{bmn}F_{cam}F_{cir}\,,\qquad C=-\frac{D}{2}\,.
\end{equation}
Thus we proved that
\begin{equation}
\label{D.9}
u_{ab}F_{acd}F_{bmn}F_{cmr}=2v_{ijb}F_{ajd}F_{bmn}F_{cam}F_{cir}\,.
\end{equation}
Using \eqn{D.9}, we can rewrite the r.h.s. of \eqn{D.4} as
\begin{equation}
\label{D.10}
-\frac{\alpha^2}{2}v_{ija}\,a_{ijardn}\,\left(J_r\otimes J_dJ_n+J_dJ_n\otimes J_r\right)\,,\qquad  a_{abcdef}=F_{adk} F_{bel} F_{cfm} F_{klm}\,.
\end{equation}
Thus \eqn{D.4} reads
\begin{equation}
\label{D.11}
2v_{ija}F_{ijb}Z_{ab}=\frac{\alpha^2}{2}v_{ija}\,a_{ijardn}\,\left(J_r\otimes J_dJ_n+J_dJ_n\otimes J_r\right)\,.
\end{equation}
Due to the contraction with a totally antisymmetric tensor we can rewrite \eqn{D.11} as
\begin{equation}
\label{D.12}
2v_{ija}F_{b[ij}Z_{a]b}=\frac{\alpha^2}{2}v_{ija}\,a_{[ija]rdn}\,\left(J_r\otimes J_dJ_n+J_dJ_n\otimes J_r\right)\,.
\end{equation}
Next, we note that
\begin{equation}
\label{D.13}
\begin{split}
&a_{[ija]rdn}=a_{ija(rdn)}\,,\\
&a_{ija(rdn)}\,\left(J_r\otimes J_dJ_n+J_dJ_n\otimes J_r\right)=a_{ijardn}\,\left(J_{(r}\otimes J_dJ_{n)}+J_{(d}J_n\otimes J_{r)}\right)\,,
\end{split}
\end{equation}
where $(.)$ denotes the sum of all permutations (see footnote \ref{permute.rule}). Using \eqn{D.12} and \eqn{D.13} we find
\begin{equation}
\label{D.14}
v_{ija}F_{b[ij}Z_{a]b}=\frac{\alpha^2}{4}v_{ija}\,a_{ijardn}\,\left(J_{(r}\otimes J_dJ_{n)}+J_{(d}J_n\otimes J_{r)}\right)\,.
\end{equation}
In addition, using \eqn{coproduct} we can easily prove that
\begin{equation}
\label{D.15}
J_{(r}\otimes J_dJ_{n)}+J_{(d}J_n\otimes J_{r)}=\frac{1}{3}\left(f\left(J_{(r}J_dJ_{n)}\right)-J_{(r}J_dJ_{n)}\otimes\mathbb{I}-\mathbb{I}\otimes J_{(r}J_dJ_{n)}\right)\,.
\end{equation}
Using \eqn{D.14} and \eqn{D.15} we find that
\begin{equation}
\label{D.16}
v_{ija}F_{b[ij}Z_{a]b}=\frac{\alpha^2}{12}v_{ija}\,a_{ijardn}\,\left(f\left(J_{(r}J_dJ_{n)}\right)-J_{(r}J_dJ_{n)}\otimes\mathbb{I}-\mathbb{I}\otimes J_{(r}J_dJ_{n)}\right)\,.
\end{equation}
Finally, we make use of the properties \eqn{D.13} to manipulate the r.h.s. of \eqn{D.16} into
\begin{equation}
\label{D.17}
\begin{split}
&v_{ija}\,a_{ijardn} W_{(rdn)}=v_{ija}\,a_{ija(rdn)} W_{rdn}=v_{ija}\,a_{[ija]rdn} W_{rdn}\,,\\
&W_{rdn}=f\left(J_{r}J_dJ_{n}\right)-J_{r}J_dJ_{n}\otimes\mathbb{I}-\mathbb{I}\otimes J_{r}J_dJ_{n}\,.
\end{split}
\end{equation}
Using \eqn{D.17} we can write \eqn{D.16} as
\begin{equation}
\label{D.18}
v_{ija}F_{b[ij}Z_{a]b}=v_{ija}\times\frac{\alpha^2}{12}\,a_{[ija]rdn} W_{rdn}\,.
\end{equation}
Since the latter holds for every antisymmetric tensor $v_{ija}$ we conclude that
\begin{equation}
\label{D.19}
\begin{split}
&F_{b[ij}Z_{a]b}=\frac{\alpha^2}{12}\,a_{[ija]rdn} W_{rdn}\Longrightarrow\\
&F_{b[ij}[Q_{a]},Q_b]=\frac{\alpha^2}{12}\,a_{[ija]rdn}\,J_rJ_dJ_n=\frac{\alpha^2}{12}\,a_{ijardn}\,J_{(r}J_dJ_{n)}\,.
\end{split}
\end{equation}
Expanding the antisymmetric part on the l.h.s. of \eqn{D.19} we find the first Serre relation \eqn{Serre}, namely
\begin{equation}
\label{D.20}
F_{mij}[Q_{k},Q_m]+F_{mki}[Q_{j},Q_m] +F_{mjk}[Q_{i},Q_m]=\frac{\alpha^2}{24}\,a_{ijkrdn}\,J_{(r}J_dJ_{n)}\,,
\end{equation}
where we used that $F_{a[bc]}=2F_{abc}.$

\subsection{The second Serre relation}

\label{Serre.II}

Applying the Jacobi identity we can easily prove that
\be
\label{C.1}
[[Q_c,Q_d],Q_e]=[[Q_c,Q_e],Q_d]-[[Q_d,Q_e],Q_c]\,,\quad F_{crk}F_{rde}-F_{drk}F_{rce}=F_{cdr}F_{rek}\,.
\ee
Using the first of \eqn{C.1} we can prove that
\ba
\label{C.2}
&&2\left(L_{ab|cd}+L_{bc|ad}+L_{ca|bd}\right)=\\
&&F_{e[ab}[[Q_{c]},Q_e],Q_d]-F_{e[ab}[[Q_{d]},Q_e],Q_c]+F_{e[cd}[[Q_{a]},Q_e],Q_b]-F_{e[cd}[[Q_{b]},Q_e],Q_a],\nonumber
\ea
where $L_{ab|cd}$ denotes the l.h.s. of \eqn{Serre2}. Using \eqn{Serre}, the second relation in \eqn{Lie} and the second equation in \eqn{C.1}, we can rewrite the r.h.s. of \eqn{C.2} as
\be
\label{C.3}
2(R_{ab|cd}+R_{bc|ad}+R_{ca|bd})\,,
\ee
where $R_{ab|cd}$ denotes the r.h.s. of \eqn{Serre2}. Combining \eqn{C.2} and \eqn{C.3} we find
\be
\label{C.4}
L_{ab|cd}+L_{bc|ad}+L_{ca|bd}=R_{ab|cd}+R_{bc|ad}+R_{ca|bd}\,.
\ee
The solution to this equation is the second Serre relation \eqn{Serre2}. In fact, we may suppose it is not, by assuming that there exists another choice of $X_{ab|cd}$ such that
\be
\label{C.5}
L_{ab|cd}=R_{ab|cd}+X_{ab|cd}\,,
\ee
where $X_{ab|cd}$ is such that
\be
\label{C.6}
X_{ab|cd}=-X_{ba|cd}=-X_{ab|dc}=X_{cd|ab}\
\ee
and
\be
\label{C.7}
X_{[ab|c]d}=0\,,\qquad X_{a[b|cd]}=0\,.
\ee
Applying the Jacobi identity on \eqn{C.5} and using \eqn{C.6} we find
\be
F_{b[mn}X_{q]b|[r\underline{d}}F_{st]d}=0\Longrightarrow X_{ab|cd}=F_{abe} Y_{e|cd}+F_{cde} Y_{e|ab}\,,\qquad Y_{a|bc}=-Y_{a|cb}\,,
\ee
where $\underline{d}$ is excluded from the anti-symmetrization.
Using these relations and \eqn{C.7}, we find
\be
Y_{a|bc}\sim F_{abc}\Longrightarrow X_{ab|cd}=\varepsilon\, F_{abe}F_{cde}\,,
\ee
where $\varepsilon$ is an arbitrary constant. Thus \eqn{C.5} reads
\be
L_{ab|cd}=R_{ab|cd}+\varepsilon\, F_{abe}F_{cde}\,.
\ee
Contracting the latter with $F_{ab\ell}F_{cd\ell}$ and using the Jacobi identity on commutators, we find
\be
0=0+\varepsilon\, c^2_G\, \mbox{dim}({\bf g})\Longrightarrow \varepsilon=0\,,
\ee
where $F_{acd}F_{bcd}=c_{\bf g}\,\d_{ab}\,, a=1,2,\dots,\mbox{dim}({\bf g}).$ This completes the proof of the redundancy of the Second Serre relation \eqn{Serre2}.

\section{The triple integral}
\label{triple}
The scope of this Appendix is to simplify \eqn{Poisson.y}.
Let us first define the triple integral as
\be
\label{B.1}
I_{ce\ell}=I_{ec\ell}=\int \mathrm{d}^3\s_{123}\,\varepsilon_{{13}}\varepsilon_{{32}}{\cal I}_0^c(\s_1){\cal I}_0^e(\s_2){\cal I}_0^\ell(\s_3)\,,
\ee
where $\mathrm{d}^3\s_{123}$ stands for $d\s_1d\s_2d\s_3$.
This can be rewritten as follows:\footnote{For manipulations of similar integrals, see \cite{Abdalla:1986xb}.}
\be
\label{B.2}
I_{ce\ell}=-\frac12\int \mathrm{d}^3\s_{123}dw\,\varepsilon_{{31}}{\cal I}_0^c(\s_1)\varepsilon_{{32}}{\cal I}_0^e(\s_2)\partial_{\s_3}\left(\varepsilon_{3w}{\cal I}_0^\ell(w)\right)\,.
\ee
Integrating by parts we can easily prove that
\ba
\label{B.3}
&&I_{ce\ell}+I_{\ell ce}+I_{e\ell c}=-Q_0^{c}Q_0^{e}Q_0^{\ell}\Longrightarrow\\
&&\left(f_{acd}f_{bre}f_{dr\ell}+f_{a\ell d}f_{brc}f_{dre}+f_{aed}f_{br\ell}f_{drc}\right)I_{ce\ell}=-f_{acd}f_{bre}f_{dr\ell}\nonumber\,
Q_0^{c}Q_0^{e}Q_0^{\ell}\,.
\ea
This formula could equivalently be found through the identity
\be
\label{B.4}
\varepsilon_{{13}}\varepsilon_{{32}}+\varepsilon_{{21}}\varepsilon_{{13}}+\varepsilon_{{32}}\varepsilon_{{21}}=-1\,.
\ee
Using the Jacobi identity we can prove that
\be
\begin{split}
\label{B.5}
&f_{a\ell d}f_{dre}f_{brc}=f_{acd}f_{bre}f_{dr\ell}+f_{aed}f_{brc}f_{dr\ell}-f_{acd}f_{br\ell}f_{dre}-f_{abr}f_{rcd}f_{d\ell e}\,,\\
&f_{aed}f_{br\ell}f_{drc}=f_{acd}f_{bre}f_{dr\ell}+f_{aed}f_{brc}f_{dr\ell}-f_{a\ell d}f_{bre}f_{drc}-f_{abr}f_{red}f_{d\ell c}\,.
\end{split}
\ee
Using the latter we can rewrite \eqn{B.3} as
\be
\begin{split}
\label{B.6}
&\left(3f_{acd}f_{bre}f_{dr\ell}+2f_{aed}f_{brc}f_{dr\ell}-f_{acd}f_{br\ell}f_{dre}-f_{a\ell d}f_{bre}f_{drc}+\right.\\
&\left.\phantom{xxxx}f_{abr}\left(f_{drc}f_{de\ell}-f_{dre}f_{d\ell c}\right)\right)I_{ce\ell}=-f_{acd}f_{bre}f_{dr\ell}\,
Q_0^{c}Q_0^{e}Q_0^{\ell}\,.
\end{split}
\ee
Using \eqn{B.4} we can prove that
\be
\label{B.7}
\left(2f_{aed}f_{brc}f_{dr\ell}-f_{acd}f_{br\ell}f_{dre}-f_{a\ell d}f_{bre}f_{drc}\right)\,I_{ce\ell}=f_{acd}f_{bre}f_{dr\ell}\left(3I_{ce\ell}+Q_0^{c}Q_0^{e}Q_0^{\ell}\right)\,.
\ee
Combining \eqn{B.6} and \eqn{B.7} we find
\be
\label{B.9}
f_{acd}f_{bre}f_{dr\ell}I_{ce\ell}=-\frac13\,f_{acd}f_{bre}f_{dr\ell}\,Q_0^{c}Q_0^{e}Q_0^{\ell}-\frac13\,f_{abr}f_{rcd}f_{de\ell}\,I_{ce\ell}\,.
\ee

\section{Serre structure constants}
\label{ident}

In this Appendix we prove \eqn{fiden}. In order to do so, we use the equivalent rewriting
\be
\label{A.1}
\frac12\,f_{d[ab}f^3_{c]d}=3f_{ai}f_{bj}f_{ck}\,f_{ijk}\,,\qquad f_{ij}:= f_{ijk}\,Q_0^{k}\,.
\ee
We start from the r.h.s. and use the Jacobi identity to rewrite ${ f_{ck}f_{ijk}}$ as
\be
\label{A.2}
f_{ck}f_{ijk}=-f_{cik}f_{jk}-f_{jck}f_{ik}\,.
\ee
Then we make use again of the Jacobi identity to deduce the following rewritings:
\be
\label{A.3}
f_{ai}f_{cik}=f_{cai}f_{ik}-f_{iak}f_{ci}\,,\qquad  f_{bj}f_{jck}=f_{bcj}f_{jk}+f_{jbk}f_{cj}\,.
\ee
Using \eqn{A.2} and \eqn{A.3}, we can rewrite the r.h.s. of \eqn{A.1} as
\be
\label{A.4}
f_{ai}f_{bj}f_{ck}f_{ijk}=f^3_{aj}f_{bcj}+f^3_{bj}f_{caj}+f^2_{bj}f_{ci}f_{iaj}-f^2_{aj}f_{ci}f_{ibj}\,,
\ee
or, equivalently, as
\ba
\label{A.5}
&&f_{ai}f_{bj}f_{ck}f_{ijk}=f^3_{bj}f_{caj}+f^3_{cj}f_{abj}+f^2_{cj}f_{ai}f_{ibj}-f^2_{bj}f_{ai}f_{icj}\,,\\
\label{A.6}
&&f_{ai}f_{bj}f_{ck}f_{ijk}=f^3_{cj}f_{abj}+f^3_{aj}f_{bcj}+f^2_{aj}f_{bi}f_{icj}-f^2_{cj}f_{bi}f_{iaj}\,.
\ea
Adding \eqn{A.4}, \eqn{A.5} and \eqn{A.6} together we find
\begin{eqnarray}
\label{A.7}
&&3f_{ai}f_{bj}f_{ck}f_{ijk}=\\
&&f_{d[ab}f^3_{c]d}-f_{aj}^2\left(f_{ci}f_{ibj}-f_{bi}f_{icj}\right)-f_{bj}^2\left(f_{ai}f_{icj}-f_{ci}f_{iaj}\right)-
f_{cj}^2\left(f_{bi}f_{iaj}-f_{ai}f_{ibj}\right) .\nonumber
\end{eqnarray}
Using the Jacobi identity we can rewrite the terms in parentheses as
\be
\label{A.8}
f_{ci}f_{ibj}-f_{bi}f_{icj}=f_{jd}f_{dbc}\,,\quad f_{ai}f_{icj}-f_{ci}f_{iaj}=f_{jd}f_{dca}\,,\quad f_{bi}f_{iaj}-f_{ai}f_{ibj}=f_{jd}f_{dab}\,.
\ee
Using \eqn{A.8}, we can rewrite \eqn{A.7} as
\be
\label{A.9}
3f_{ai}f_{bj}f_{ck}f_{ijk}=f_{d[ab}f^3_{c]d}-\frac12\,f_{d[ab}f^3_{c]d}=\frac12\,f_{d[ab}f^3_{c]d}\,,
\ee
which completes the proof of \eqn{A.1} or equivalently \eqn{fiden}.

\section{Maillet brackets and the first Serre relation}
\label{Maillet.expand}

Let us first consider the l.h.s. of (\ref{Mbrac}) at the order ${\cal{O}}(\nu'^2 \mu'^2)$ of the expansion of the monodromy matrix \eqref{expansionn}. Specifically, if we define
\begin{eqnarray}
\label{charges.rescale}
g_1 :=  \frac{\nu'}{1-\rho^2}, \quad g_2 :=  \frac{\mu'}{1-\rho^2 },\quad
q=\frac{Q_0}{2}\,,\quad \widehat q=\frac{Q_1}{2}\,,
\end{eqnarray}
we get
\begin{eqnarray}
4 \, g_1^2 \, g_2^2 \, \{\big(\widehat{q} - \rho \, q + q^2\big),\big(\widehat{q} - \rho \, q + q^2\big)\}.
\end{eqnarray}
Expanding the Poisson brackets and using \eqref{Maillet.lowest} written in terms of $q,\widehat q$, {\it i.e.}
\begin{eqnarray}
\label{appendix.zeros}
\{q_a, q_b\} \, = \Delta f_{abc} q_c, \quad \{q_a, \widehat{q}_b\} \, = \Delta f_{abc} \widehat{q}_c,\quad
\Delta= \, - e^2 \, \left(\rho^2 + 1 + x(1- \rho^2)\right),
\end{eqnarray}
we obtain
\begin{eqnarray}
&&4 \, g_1^2 \, g_2^2 \, \bigg[ \{\widehat{q}_a, \widehat{q}_b\} \, T_a \otimes T_b \, + \, 2 \, \rho \, \Delta \, f_{bac} \, \widehat{q}_c \, T_a \otimes T_b \, - \, \Delta \, f_{bad} \, \widehat{q}_d \, q_c \, T_a \otimes \{T_b, T_c\}  \nonumber\\
&&+\rho^2 \, \Delta \, f_{abc} \, q_c \, T_a \otimes T_b \, - \, \rho \, \Delta \, f_{abd} \, q_d \, q_c \, T_a \otimes \{T_b, T_c\} \, + \, \Delta \, f_{acd} \,   \widehat{q}_d \, q_b \, \{T_a, T_b\} \otimes T_c \nonumber\\
&&- \rho \, \Delta \, f_{bcd} \, q_a \, q_d \, \{T_a, T_b\} \otimes T_c \, + \, \Delta \, f_{bce} \, q_a \, q_e \, q_ d \, \{T_a, T_b\} \otimes \{T_c, T_d\} \bigg],
\end{eqnarray}
where $\{T_a, T_b\} := T_a T_b + T_b T_a$.

We now need to consider the r.h.s. of (\ref{Mbrac}) at the same order. Let us define
\begin{eqnarray}
q_1 \, := \, q \otimes \mathbb{I}, \qquad q_2 \, := \, \mathbb{I} \otimes q,
\end{eqnarray}
as in footnote \ref{tensor.product}.
There are several contributions, which we list here below:

\begin{itemize}
\item both $r(\nu', \mu')$ and $s(\nu', \mu')$ are taken at the order ${\cal{O}}(\nu' \mu')$, which means that both ${M}(\nu')$ and ${M}(\mu')$ are taken at the linear order. This term contributes
\begin{eqnarray}
16 \, e^2 \, \rho \, g_1^2 \, g_2^2 \bigg(- \big[\Pi, q_1 \, q_2] \, + \, q_1 \, \Pi \, q_2 \, - \, q_2 \, \Pi \, q_1 \bigg);
\end{eqnarray}
\item $r(\nu', \mu')$ is taken at the order ${\cal{O}}(\nu' \mu'^{\, 0})$, which means that ${M}(\nu')$ is taken at the linear and ${M}(\mu')$ at the quadratic order. No $s(\nu', \mu')$ contribution is present at this order. We get
\begin{eqnarray}
- 8 \, \Delta \, g_1^2 \, g_2^2 \, \big[\Pi, q_1 \, (\widehat{q}_2 - \rho \, q_2 + q_2^2)\big];
\end{eqnarray}
\item $r_-(\nu', \mu')$ is taken at the order ${\cal{O}}(\nu' \mu'^2)$, which means that ${M}(\nu')$ is taken at the linear and ${M}(\mu')$ at the zeroth order (making $r_+$ immaterial). We get
\begin{eqnarray}
8 \, e^2 \, (1 + 3 \rho^2) \, g_1^2 \, g_2^2 \, \big[\Pi, q_1 \big];
\end{eqnarray}
\item $r_+(\nu', \mu')$ is taken at the order ${\cal{O}}(\nu'^2 \mu'^{\, 0})$, which means that ${M}(\nu')$ is taken at the zeroth and ${M}(\mu')$ at the quadratic order (making $r_-$ irrelevant). We get
\begin{eqnarray}
- 16 \, e^2 \, \rho \, g_1^2 \, g_2^2 \, \big[\Pi, (\widehat{q}_2 - \rho \, q_2 + q_2^2) \big];
\end{eqnarray}
\item $r_+(\nu', \mu')$ is taken at the order ${\cal{O}}(\nu'^2 \mu')$, which means that ${M}(\nu')$ is taken at the zeroth and ${M}(\mu')$ at the linear order. We will show that this term does not contribute to the final result, upon applying the procedure (\ref{oper}) we will introduce shortly.
\item $r_+(\nu', \mu')$ is taken at the order ${\cal{O}}(\nu'^2 \mu'^{\, - 1})$, which means that ${M}(\nu')$ is taken at the zeroth and ${M}(\mu')$ at the cubic order. The presence of this negative power in the expansion of $r_+$ forces us to go to the third order in the expansion of the monodromy matrix, which we will perform later on - {{} see the discussion around (\ref{oper})}.

Putting all the terms together and performing a few manipulations, we get for the r.h.s. of the Poisson relations
\begin{eqnarray}
&&16 \, e^2 \, \rho \, g_1^2 \, g_2^2 \, q_a \, q_b \, \bigg(f_{acd} \, f_{bce} \, T_d \otimes T_e \, + \, \frac{\Delta}{2 \, e^2} \, f_{cbd} \, T_a \, T_c \otimes T_d - \, f_{cbe} \, T_c \otimes T_a \, T_e \nonumber\\
&&+ \, \bigg[\frac{\Delta}{2 \, e^2} \, - \, 1\bigg] \, f_{acd} \, T_c \otimes T_d \, T_b \, \bigg) \, - \, 8 \, \Delta \, g_1^2 \, g_2^2 \, q_a \, \widehat{q}_b \, (f_{cbd} \, T_a \, T_c \otimes T_d \, + \nonumber\\
&&\, f_{cad} \, T_d \otimes T_c \, T_b) \, - \, 8 \, \Delta \, g_1^2 \, g_2^2 \, q_a \, q_b \, q_d \, (f_{cbe} \, T_c \, T_a \otimes T_e \, T_d \, + \, f_{cde} \, T_c \, T_a \otimes T_b \, T_e \, \nonumber\\
&&+ \, f_{ace} \, T_e \otimes T_b \, T_d \, T_c) - \, 16 \, e^2 \, \rho \, g_1^2 \, g_2^2 \, \widehat{q}_b \, f_{cba} \, T_c \otimes T_a +\nonumber\\
&&8 \, e^2 \, (1 + 3 \rho^2) \, g_1^2 \, g_2^2 \, q_a \, \bigg[f_{abc} \, T_b \otimes T_c \, - \frac{2 \rho^2}{(1 + 3 \rho^2)} \, f_{abc} \, T_b \otimes T_c \, \bigg] \, .
\end{eqnarray}
\end{itemize}

The strategy we will now follow is to bring everything on one side of the equation, namely to calculate
\begin{eqnarray}
\label{apper}
\frac{\mbox{l.h.s.} - \mbox{r.h.s.}}{g_1^2 \, g_2^2},
\end{eqnarray}
and to act upon it with the following operation:
\begin{eqnarray}
\label{oper}
\frac{\Delta}{2} \, f_{\delta [\alpha \beta} \, \tr \, \Big( T_{\gamma]} \otimes T_\delta \, \circ \, \Big),
\end{eqnarray}
where the three indices $\alpha, \beta$ and $\gamma$ are totally antisymmetrized (without the $\frac{1}{6}$ factor).
Upon performing the operation (\ref{oper}), and by using the Jacobi identity, the very first term of the l.h.s.
as contributing to (\ref{apper}), namely $4 \{\widehat{q}_a, \widehat{q}_b\} \, T_a \otimes T_b$, can be seen to coincide with
\begin{eqnarray}
4\{\widehat{q}_\alpha, \{ \widehat{q}_\beta, q_\gamma\}\} \, - \, 4\{q_\alpha, \{ \widehat{q}_\beta, \widehat{q}_\gamma\}\},
\end{eqnarray}
which is the desired combination appearing in the Serre relations. It is therefore a matter of analyzing all the other terms after this operation is performed.
One thing to notice is that anything looking like
\begin{eqnarray}
\label{form}
f_{\gamma \delta b} \Omega^b
\end{eqnarray}
will vanish upon this operation, as can be seen by using the Jacobi identity. This is the reason why
the contribution of $r_+(\nu', \mu')$ taken at the order ${\cal{O}}(\nu'^2 \mu')$ is absent,
as we commented earlier, since it is precisely of the form (\ref{form}).
Disregarding this type of terms as irrelevant to the final result, we combine the remaining terms in (\ref{apper}) and
perform quite extensive manipulations and simplifications. Performing then the operation (\ref{oper})
on the result of this simplification produces some terms that we will call {\it unwanted}, since they do
not look like the standard terms appearing in the Serre relations, and some that we call {\it wanted}, since they have the desired form.
\subsection{Unwanted terms}
Let us begin with the unwanted terms. They come in two fashions:
\begin{itemize}
\item We get a quadratic contribution with level zero charges, specifically
\begin{eqnarray}
\label{primo}
- 8 \, \Delta \, \rho \, q_a \, q_b \, f_{\gamma e b} \, \tr \Big( T_\delta \{T_e, T_a\}\Big) \, + \,
\mbox{contrib. $r_+(\nu', \mu')$ at  ${\cal{O}}(\nu'^2 \mu'^{\, - 1})$};
\end{eqnarray}
\item We get a quadratic term with level zero and one charges, specifically
\be
\begin{split}
\label{second}
&4 \, \Delta \, q_a \, \widehat{q}_b \, \bigg[- f_{e \gamma b} \, \tr \Big( T_\delta \{T_e, T_a\}\Big) \, - \,
f_{e \gamma a} \tr \Big( T_\delta \{T_e, T_b\}\Big) \bigg] \, + \\
&\mbox{contrib. from $r_+(\nu', \mu')$ at  ${\cal{O}}(\nu'^2 \mu'^{\, - 1})$}.
\end{split}
\ee
\end{itemize}

The respective first terms in (\ref{primo}) and (\ref{second}) vanish upon the operation (\ref{oper}).
In order to see this, one needs to proceed in steps. Let us consider the first unwanted term. The first step consists of
repeatedly using
\begin{eqnarray}
f_{abc} = \tr \Big(T_a [T_b, T_c]\Big)\,,
\end{eqnarray}
to re-write the first term in (\ref{primo}), after the action of (\ref{oper}), as (indicating only the matrix part)
\begin{eqnarray}
\label{struc}
- \frac{1}{2} f_{\delta [\alpha \beta} \, f_{\gamma] e b} \, \tr \Big( T_\delta \{T_e, T_a\}\Big) \, =  \, \tr \Big(\{T_e, T_a\} \, [T_\alpha, T_\beta] \Big) \, \tr \Big(T_e \, [T_\gamma, T_b] \Big) \, + \, \mbox{"2"},
\end{eqnarray}
where "2" means that we have to add the other two cyclic permutations $\beta \gamma \alpha$ and $\gamma \alpha \beta$
of the same structure on the l.h.s. of (\ref{struc}). At this point, it is convenient to open up the anti-commutator,
involving the generator $T_e$, and move $T_e$ close to the other trace by using cyclicity of the trace.
When this is done, it produces two terms, in each of which one recognizes a structure of the type
\begin{eqnarray}
\tr \Big(x \, T_e\Big) \, \tr \Big(T_e \, y \Big), \qquad y \in \mbox{Lie algebra}.
\end{eqnarray}
The fact that $y$ is Lie-algebra valued makes it possible to fuse the traces producing $\tr (x \, y)$. For the
matrix part of the first unwanted term, we are therefore left with
\begin{eqnarray}
\tr \Big(T_a \, [T_\alpha, T_\beta] \, [T_\gamma, T_b]\Big) \, + \, \tr \Big([T_\alpha, T_\beta] \, T_a \, [T_\gamma, T_b]\Big) \, + \, \mbox{"2"}.
\end{eqnarray}
Adding the "2" explicitly to this term, symmetrizing $a \leftrightarrow b$ given the $q_a \, q_b $ in front of
the first term in (\ref{primo}) and eventually expanding all the (anti-)commutators explicitly, one sees that
all terms cancel and the total contribution vanishes. We will calculate the contribution from $r_+(\nu', \mu')$ at  ${\cal{O}}(\nu'^2 \mu'^{\, - 1})$ later on.

With regards to the first term in (\ref{second}), let us re-write it as
\begin{eqnarray}
4 \, \Delta (q_a \, \widehat{q}_b + q_b \, \widehat{q}_a) f_{\gamma e b} \, \tr \Big( T_\delta \{T_e, T_a\}\Big).
\end{eqnarray}
From this we see that, due to the $a \leftrightarrow b$ symmetry of the pre-factor, perfectly analogous
considerations apply as for the term we have just shown to vanish. Once again, we will study the contribution
from $r_+(\nu', \mu')$ at  ${\cal{O}}(\nu'^2 \mu'^{\, - 1})$ later on.

\subsection{Third order of the monodromy}

We have seen that, to be able to calculate the two {\it unwanted} terms left-over from (\ref{primo}) and (\ref{second}),
and also the related contribution to the {\it wanted} terms, we need the monodromy matrix up to the third order in the spectral parameter.
We will derive this term in the expansion in this section.

From {section \ref{Lax}} we have learned that the monodromy matrix $M$ admits an expansion (adapted to the parameter $g$ used in this section)
{
\begin{eqnarray}
M(g) \, = \, P \exp\,g\int\frac{{\cal I}_0 - \, s \, g \, I_1}{(1+\rho g)^2-g^2}\,,
\end{eqnarray}
}
where { we note that both ${\cal I}_0$ and $I_1$ appear and}
\begin{eqnarray}
s \, = \, \rho^2 - 1.
\end{eqnarray}
We need to isolate the third order term in $g$. If we recall the density \eqref{charges1} of the level-zero charge
\begin{eqnarray}
2 \, j(\sigma) \, = \, {\cal I}_0(\sigma)\,,
\end{eqnarray}
we see that the third-order term we are after reads, in compact notation,
\begin{eqnarray}
\label{M3}
&&2 \, g^3 \, (4 \rho^2 - s) \, \int j \, + \, 2 \, \rho \, s \, g^3 \, \int I_1 \, - 16 \, \rho \, g^3 \,
\int \int^\sigma j(\sigma) \, j(\sigma') \\
&&- \, 2 \, s \, g^3 \, \int \int^\sigma \Big[j(\sigma) \, I_1(\sigma') \, +  I_1(\sigma) \, j(\sigma')\Big] \,
+ \, 8 \, g^3 \, \int \int^\sigma \int^{\sigma'} \, j(\sigma) \, j(\sigma') \, j(\sigma'') \, := \, M_3. \nonumber
\end{eqnarray}
We have to put this term into a form which is ready to be used for the Serre relations, therefore we parameterize
\begin{eqnarray}
\frac{M_3}{g^3} \, = \, A \, Q_0^3 \, + \, \frac{B}{2} \, Q_0 \, \widehat{Q} \, + \, \frac{C}{2} \, \widehat{Q} \, Q_0 \,
+ \, D \, Q_0^2 \, + \, \frac{E}{2} \, \widehat{Q} \, + \, F \, Q_0 \, + \, \mbox{"Lie"},
\end{eqnarray}
where we recall that (in the notation of this section) $\widehat{Q}$ reads
\begin{eqnarray}
\widehat{Q} \, = \, \int (- 2\rho \, j \, - \, s \, I_1) \, + \, 4 \, \int \int^\sigma j(\sigma) \, j(\sigma').
\end{eqnarray}
"Lie" carries such a name because it is something which is not easily expressed in terms of $Q$ or $\widehat{Q}$,
nevertheless it is Lie-algebra valued, hence it will drop after operation (\ref{oper}). In particular, we choose
\begin{eqnarray}
&&\mbox{"Lie"} \, = \, H \, \int \int^\sigma [j(\sigma), \, j(\sigma')] \, + \, \frac{V}{2} \, \int \int^\sigma [j(\sigma), \,
I_1(\sigma')] \, + \, \frac{U}{2} \, \int \int^\sigma [I_1(\sigma), \, j(\sigma')] \, + \nonumber\\
&&N \, \int \int^\sigma \int^{\sigma'} \, [j(\sigma), \,[ j(\sigma'), \, j(\sigma'')]] \, + \, P \, \int \int_\sigma \int^{\sigma} \,
[j(\sigma), \,[ j(\sigma'), \, j(\sigma'')]] \, + \nonumber\\
&&R \, \int \int_\sigma \int_\sigma^{\sigma'} \, [j(\sigma), \,[ j(\sigma'), \, j(\sigma'')]].
\end{eqnarray}
By appropriately splitting the integration domains and taking into account the ordering of the generators, one can show that
the terms we use in our parametrization of $M_3$ are enough to reconstruct the most general integral appearing at this order.
In fact, we find in this way that they are more than sufficient, as we find a family of solutions when we try and match with (\ref{M3}):
\be
\begin{split}
&A = -\frac{8}{3}, \qquad F \, = \, \rho \, E \, + \, 2 \, (4 \rho^2 - s), \qquad G \, = \, s (E + 4 \rho) \qquad V \, = \, s \, (B - 4),  \\
&U \, = \, - s \, B, \qquad C \, = \, 4 \, - \, B, \qquad R \, = \, N \, - \, \frac{8}{3}, \qquad P \, = \, \frac{16}{3} \, - \, 2 \, B \, - \, N, \qquad  \\
&H \, = \, - E \, - 8 \, \rho, \qquad D \, = \, - E \, - 4 \, \rho,
\end{split}
\ee
from which we see that we can set
\begin{eqnarray}
B \, = \, E \, = \, N \, = \, 0
\end{eqnarray}
as a convenient choice.

The only contribution from the third order of the monodromy that can survive the operation (\ref{oper}) is then
\begin{eqnarray}
\label{unw3}
g^3 \, \Big[\frac{4}{3} \, q^3 \, + \, 4 \, \widehat{q} \, q \, - \, 4 \, \rho \, q^2\Big].
\end{eqnarray}

\begin{itemize}

\item The $q^2$ term in (\ref{unw3}) represents the contribution to the unwanted term (\ref{primo}) from the third order expansion in the monodromy.
After acting with (\ref{oper}) and performing a few manipulations on the indexes, one can see that this contribution reproduces the same structure
as the first addendum in (\ref{primo}), hence it vanishes for the same reasons.

\item The $\widehat{q} \, q$ term in (\ref{unw3}) combines with the order ${\cal{O}}(\nu'^2 \mu'^{\, -1})$ of $r_+ (\nu', \mu')$ to give
\begin{eqnarray}
- 2 \, \Delta \, g_1^2 \, g_2^2 \, \widehat{q}_b \, q_d \, \Big[f_{cbe} \, T_c \otimes T_e \, T_d \, + \, f_{cde} \, T_c \otimes T_b \, T_e \Big].
\end{eqnarray}
After acting with (\ref{oper}), one can see that this term as well reduces the same structure as the first addendum in (\ref{second}),
hence vanishing by the same token.

\end{itemize}

\subsection{Wanted terms}

We now need to calculate the contribution we do expect to appear in the first Serre relation, namely we evaluate the cubic term in the
level-zero charges in the expression (\ref{apper}). This amounts to the following - after acting with (\ref{oper}):
\be
\begin{split}
\label{sopra}
&2 \, \Delta^2 \, q_a \, q_b \, q_d \, f_{\delta [\alpha \beta} \, \Big[4 \, \tr (T_{\gamma]} \, T_a \, T_e) \, \tr (T_\delta \, T_c \, T_d) f_{ecb} \,
+ \, 2 \, f_{\gamma] e a} \, f_{ecb} \, \tr (T_\delta \, T_c \, T_d) \, + \,\\
&2 \, \tr (T_{\gamma]} \, T_a \, T_e) \, f_{ecb} \, f_{d c \delta} \, + \, f_{\gamma] e a} \, f_{ecb} \, f_{d c \delta} \, + \, 2 \, \tr (T_{\gamma]} \,
T_c \, T_a) \, \tr (T_\delta \, T_e \, T_d) \,f_{cbe} \, + \, \\
&2 \, \tr (T_{\gamma]} \, T_c \, T_a) \, \tr (T_\delta \, T_b \, T_e) \,f_{cde} \, - \, 2 \, f_{\gamma] a c} \, \tr (T_\delta \, T_b \, T_d \, T_c)\Big]
 \\
&+ \, \mbox{contribution from $r_+(\nu', \mu')$ at  ${\cal{O}}(\nu'^2 \mu'^{\, - 1})$}\ .
\end{split}
\ee
Let us start with the part in (\ref{sopra}) which is not coming from $r_+(\nu', \mu')$ at  ${\cal{O}}(\nu'^2 \mu'^{\, - 1})$.
Exploiting the total symmetry of the pre-factor $q_a \, q_b \, q_c$, after repeated use of the Jacobi identity, and by use of
reconstructing commutators inside the traces to reduce the length of the traces as much as possible, we can recast that contribution into
\begin{eqnarray}
- 4 \, \Delta^2 \, q_a \, q_b \, q_c \, f_{\delta [\alpha \beta} \, f_{\gamma] b e} \, \tr (T_\delta \, T_a \, T_e \, T_c) \, + \, 2 \,
\Delta^2 \, q_a \, q_b \, q_c \, f_{\delta [\alpha \beta} \, f_{\gamma] b r} \, f_{rce} \, f_{a e \delta}\ .
\end{eqnarray}
Before proceeding with the calculation, let us compare with the contribution from $r_+(\nu', \mu')$ at  ${\cal{O}}(\nu'^2 \mu'^{\, - 1})$
and see whether the difficult-to-handle length-four trace cancels. It does indeed, since, by performing similar manipulations,
the contribution from $r_+(\nu', \mu')$ at  ${\cal{O}}(\nu'^2 \mu'^{\, - 1})$ term results into
\begin{eqnarray}
4 \, \Delta^2 \, q_a \, q_b \, q_c \, f_{\delta [\alpha \beta} \, f_{\gamma] b e} \, \tr (T_\delta \, T_a \, T_e \, T_c) \, - \, \frac{4}{3} \,
\Delta^2 \, q_a \, q_b \, q_c \, f_{\delta [\alpha \beta} \, f_{\gamma] b r} \, f_{rce} \, f_{a e \delta}\ .
\end{eqnarray}

We are then left with the two purely structure-constant contribution, which, by repeated use of the Jacobi identity, can be
combined and manipulated into
\begin{eqnarray}
- 4 \, \Delta^2 \, q_a \, q_b \, q_c \, f_{\alpha a e} \, f_{\beta b d} \, f_{\gamma c r} \, f_{edr}.
\end{eqnarray}
We now recall that this contributes to "l.h.s. - r.h.s." of the Poisson brackets, hence it changes sign when brought back to the r.h.s., giving
\begin{eqnarray}
\label{final}
\{\widehat{q}_\alpha, \{ \widehat{q}_\beta, q_\gamma\}\} \, - \, \{q_\alpha, \{ \widehat{q}_\beta, \widehat{q}_\gamma\}\} \, = \, \frac{1}{6} \, \Delta^2 \,
q_{(a} q_b q_{c)} \, f_{\alpha a e} \, f_{\beta b d} \, f_{\gamma c r} \, f_{edr}\ .
\end{eqnarray}
Using the Jacobi identity, eqs. \eqref{charges.rescale}, \eqref{appendix.zeros} and the definition of $\Delta$ (\ref{deltadef}),
we ultimately re-obtain the first Serre relation \eqref{SerreI}.


\begin{thebibliography}{1}

\bibitem{Sfetsos:2013wia}
  K.~Sfetsos,
  {\it Integrable interpolations: From exact CFTs to non-Abelian T-duals,}
  Nucl.\ Phys.\ B {\bf 880} (2014) 225,
 \href{http://arxiv.org/abs/1312.4560}{arXiv:1312.4560 [hep-th].}

\bibitem{Balog:1993es}
  J.~Balog, P.~Forgacs, Z.~Horvath and L.~Palla,
 {\it A New family of su(2) symmetric integrable sigma models,}
  Phys. Lett. {\bf B324} (1994) 403, \href{http://arxiv.org/abs/hep-th/9307030}{hep-th/9307030}.

  \bibitem{Evans:1994hi}
  J.~M.~Evans and T.~J.~Hollowood,
 {{\it Integrable theories that are asymptotically CFT}},
  Nucl.\ Phys.\ B {\bf 438} (1995) 469
  \href{http://arxiv.org/abs/hep-th/9407113}{hep-th/9407113.}

  \bibitem{Hollowood:2014rla}
  T.~J.~Hollowood, J.~L.~Miramontes and D.~M.~Schmidtt,
 {\it Integrable Deformations of Strings on Symmetric Spaces,}
 \href{http://arxiv.org/abs/1407.2840}{arXiv:1407.2840 [hep-th].}

\bibitem{Buscher:1987sk}
  T.~H.~Buscher,
  {\it A Symmetry of the String Background Field Equations,}
  \href{http://www.sciencedirect.com/science/article/pii/0370269387907696}{Phys. Lett. {\bf B194} (1987) 59} and
{\it Path Integral Derivation of Quantum Duality in Nonlinear Sigma Models,}
  \href{http://www.sciencedirect.com/science/article/pii/0370269388906028}{Phys. Lett. {\bf B201} (1988) 466.}

\bibitem{Fridling:1983ha}
  B.~E.~Fridling and A.~Jevicki,
  {\it Dual Representations and Ultraviolet Divergences in Nonlinear $\sigma$-Models,}
\href{http://www.sciencedirect.com/science/article/pii/0370269384909870}{Phys. Lett. {\bf B134} (1984) 70}.

\bibitem{Fradkin:1984ai}
  E.~S.~Fradkin and A.~A.~Tseytlin,
  {\it Quantum Equivalence of Dual Field Theories,}
 \href{http://www.sciencedirect.com/science/article/pii/0003491685902258}{Annals Phys. {\bf 162} (1985) 31}.

\bibitem{delaOssa:1992vc}
  X.~C.~de la Ossa and F.~Quevedo,
  {\it Duality symmetries from nonAbelian isometries in string theory,}
  Nucl. Phys. {\bf B403} (1993) 377,
  \href{http://arxiv.org/abs/hep-th/9210021}{hep-th/9210021}.


\bibitem{Curtright:1994be}
  T.~Curtright and C.~K.~Zachos,
  {\it Currents, charges, and canonical structure of pseudodual chiral models,}
  Phys. Rev. {\bf D49} (1994) 5408,
  \href{http://arxiv.org/abs/hep-th/9401006}{hep-th/9401006}.

\bibitem{Lozano:1995jx}
  Y.~Lozano,
  {\it Non-Abelian duality and canonical transformations},
  Phys. Lett. {\bf B355} (1995) 165,
 \href{http://arxiv.org/abs/hep-th/9503045}{hep-th/9503045}.

\bibitem{Sfetsos:1996pm}
  K.~Sfetsos,
  {\it Non-Abelian duality, parafermions and supersymmetry},
  Phys. Rev. {\bf D54} (1996) 1682,
  \href{http://arxiv.org/abs/hep-th/9602179}{hep-th/9602179}.

\bibitem{Sfetsos:2010uq}
  K.~Sfetsos and D.~C.~Thompson,
  {\it On non-abelian T-dual geometries with Ramond fluxes},
  Nucl. Phys. {\bf B846} (2011) 21, \href{http://arxiv.org/abs/arXiv:1012.1320}{arXiv:1012.1320 [hep-th]}.

\bibitem{Itsios:2013wd}
  G.~Itsios, C.~N\'u\~{n}ez, K.~Sfetsos and D.~C.~Thompson,
  {\it Non-Abelian T-duality and the AdS/CFT correspondence: new $\cN=1$ backgrounds},
  Nucl. Phys. {\bf B873} (2013) 1, \href{http://arxiv.org/abs/arXiv:1301.6755}{arXiv:1301.6755 [hep-th]}.

\bibitem{Barranco:2013fza}
  A.~Barranco, J.~Gaillard, N.~T.~Macpherson, C.~N\'u\~{n}ez and D.~C.~Thompson,
  {\it G-structures and Flavouring non-Abelian T-duality},
  JHEP {\bf 1308} (2013) 018,
 \href{http://arxiv.org/abs/arXiv:1305.7229}{arXiv:1305.7229 [hep-th]}.

\bibitem{Lozano:2013oma}
  Y.~Lozano, E.~\'O. Colg\'ain and D.~Rodr\'iguez-G\'omez,
  {\it Hints of 5d Fixed Point Theories from Non-Abelian T-duality},
  JHEP {\bf 1405} (2014) 009,
  \href{http://arxiv.org/abs/arXiv:1311.4842}{arXiv:1311.4842 [hep-th]}.

 \bibitem{Itsios:2014lca}
  G.~Itsios, K.~Sfetsos and K.~Siampos,
  {\it The all-loop non-Abelian Thirring model and its RG flow},
  Phys. Lett. {\bf B733} (2014) 265,
 \href{http://arxiv.org/abs/arXiv:1404.3748}{arXiv:1404.3748 [hep-th].}

\bibitem{Sfetsos:2014jfa}
  K.~Sfetsos and K.~Siampos,
  {\it Gauged WZW-type theories and the all-loop anisotropic non-Abelian Thirring model},
  Nucl. Phys. {\bf B885}, 583,
  \href{http://arxiv.org/abs/arXiv:1405.7803}{arXiv:1405.7803 [hep-th].}

\bibitem{Drinfeld:1985rx}
  V.~G.~Drinfeld,
 {\it Hopf algebras and the quantum Yang-Baxter equation},
  Sov.\ Math.\ Dokl.\  {\bf 32} (1985) 254
   [Dokl.\ Akad.\ Nauk Ser.\ Fiz.\  {\bf 283} (1985) 1060], 
   Michio Jimbo, ed. \href{http://www.worldscientific.com/worldscibooks/10.1142/1021}
   {{\it Yang-Baxter Equation in Integrable Systems}}. Advanced Series in Mathematical Physics Vol. 10. World Scientific, p. 264, 1990.

\bibitem{Bernard:1992ya}
  D.~Bernard,
  {\it An Introduction to Yangian Symmetries},
  Int. J. Mod. Phys. {\bf B7} (1993) 3517,
  \href{http://arxiv.org/abs/hep-th/9211133}{hep-th/9211133}.

\bibitem{MacKay:2004tc}
  N.~J.~MacKay,
 {\it Introduction to Yangian symmetry in integrable field theory,}
  Int.\ J.\ Mod.\ Phys.\ A {\bf 20} (2005) 7189,
  \href{http://arxiv.org/abs/hep-th/0409183}{hep-th/0409183.}

\bibitem{Torrielli:2010kq}
  A.~Torrielli,
  {\it Review of AdS/CFT Integrability, Chapter VI.2: Yangian Algebra},
  Lett. Math. Phys. {\bf 99} (2012) 547,
 \href{http://arxiv.org/abs/arXiv:1012.4005}{arXiv:1012.4005 [hep-th]}.

\bibitem{MacKay:1992he}
  N.~J.~MacKay,
  {\it On the classical origins of Yangian symmetry in integrable field theory},
  \href{http://www.sciencedirect.com/science/article/pii/037026939290280H}{Phys. Lett. {\bf B281} (1992) 90}
  \href{http://www.sciencedirect.com/science/article/pii/037026939391310J}{[Erratum-ibid. {\bf B308} (1993) 444].}

\bibitem{Maillet2}
  J.~M.~Maillet,
 {\it New Integrable Canonical Structures in Two-dimensional Models},
 \href{http://www.sciencedirect.com/science/article/pii/0550321386903652}{Nucl. Phys. {\bf B269} (1986) 54.}

\bibitem{Maillet}
  J.~M.~Maillet,
  {\it Hamiltonian Structures for Integrable Classical Theories From Graded Kac-moody Algebras},
 \href{http://www.sciencedirect.com/science/article/pii/037026938691289X}{Phys. Lett. {\bf B167} (1986) 401.}

\bibitem{Luscher:1977rq}
  M.~Luscher and K.~Pohlmeyer,
  {\it Scattering of Massless Lumps and Nonlocal Charges in the Two-Dimensional Classical Nonlinear Sigma Model},
  \href{http://www.sciencedirect.com/science/article/pii/0550321378900494}{Nucl. Phys. {\bf B137} (1978) 46.}

\bibitem{FTbook}
L.~D. Faddeev and L.~A. Takhtajan, {\it Hamiltonian methods in the theory of solitons}, \href{http://link.springer.com/book/10.1007%2F978-3-540-69969-9}{Springer,
Berlin 1987.}

\bibitem{Evgeny}
  E.~K.~Sklyanin,
  {\it Quantum version of the method of inverse scattering problem},
\href{http://link.springer.com/article/10.1007\%2FBF01091462} {J. Sov. Math. {\bf 19} (1982) 1546}
   [Zap.\ Nauchn.\ Semin.\  {\bf 95} (1980) 55].

\bibitem{Rajeev:1988hq}
  S.~G.~Rajeev,
  {\it Nonabelian Bosonization Without Wess-zumino Terms. 1. New Current Algebra},
\href{http://www.sciencedirect.com/science/article/pii/0370269389915281}{Phys. Lett. {\bf B217} (1989) 123}.

\bibitem{Abdalla:1986xb}
  E.~Abdalla, M.~C.~B.~Abdalla and M.~Forger,
  {\it Exact $S$-Matrices for Anomaly Free Nonlinear $\sigma$-Models on Symmetric Spaces},
  \href{http://www.sciencedirect.com/science/article/pii/0550321388900259}{Nucl. Phys. {\bf B297} (1988) 374}.

\bibitem{Cherednik:1981df}
  I.~V.~Cherednik,
  {\it Relativistically Invariant Quasiclassical Limits of Integrable Two-dimensional Quantum Models},
  \href{http://link.springer.com/article/10.1007\%2FBF01086395#page-1}{Theor.\ Math.\ Phys.\  {\bf 47} (1981) 422}
   [Teor.\ Mat.\ Fiz.\  {\bf 47} (1981) 225].

\bibitem{Hlavaty}
L.~Hlavat\'y,
{\it On the Lax formulation of generalized SU(2) principal models},\hfill\break
\href{http://www.sciencedirect.com/science/article/pii/S0375960100003534}{Phys. Lett. {\bf A271} (2000) 207.}

\bibitem{Mohammedi:2008vd}
  N.~Mohammedi,
  {\it On the geometry of classically integrable two-dimensional non-linear sigma models},
  Nucl. Phys. {\bf B839} (2010) 420,
\href{http://arxiv.org/abs/arXiv:0806.0550}{arXiv:0806.0550 [hep-th]}.

 { 
  \bibitem{Kawaguchi:2011mz}
  I.~Kawaguchi, D.~Orlando and K.~Yoshida,
  {\it Yangian symmetry in deformed WZNW models on squashed spheres},
  Phys.\ Lett.\ B {\bf 701} (2011) 475,
  \href{http://arxiv.org/abs/1104.0738}{arXiv:1104.0738 [hep-th].}
  
  \bibitem{Kawaguchi:2013gma}
  I.~Kawaguchi and K.~Yoshida,
  {\it A deformation of quantum affine algebra in squashed Wess-Zumino-Novikov-Witten models},
  J.\ Math.\ Phys.\  {\bf 55} (2014) 062302,
  \href{http://arxiv.org/abs/1311.4696}{arXiv:1311.4696 [hep-th].}
}


\bibitem{Delduc:2013fga}
  F.~Delduc, M.~Magro and B.~Vicedo,
  {\it On classical $q$-deformations of integrable sigma-models,}
  JHEP {\bf 1311} (2013) 192,
 \href{http://arxiv.org/abs/1308.3581}{arXiv:1308.3581 [hep-th].}


\bibitem{Delduc:2013qra}
  F.~Delduc, M.~Magro and B.~Vicedo,
  {\it An integrable deformation of the $AdS_5 \times S^5$ superstring action,}
  Phys.\ Rev.\ Lett.\  {\bf 112} (2014) 051601,
 \href{http://arxiv.org/abs/1309.5850}{arXiv:1309.5850 [hep-th].}

\bibitem{Delduc:2014kha}
  F.~Delduc, M.~Magro and B.~Vicedo,
 {\it Derivation of the action and symmetries of the q-deformed $AdS_5 \times S^5$ superstring,}
 \href{http://arxiv.org/abs/1406.6286}{arXiv:1406.6286 [hep-th].}

\bibitem{Klimcik:2002zj}
  C.~Klimcik,
  {\it Yang-Baxter sigma-models and dS/AdS T duality,}
  JHEP {\bf 0212} (2002) 051,
 \href{http://arxiv.org/abs/hep-th/0210095}{hep-th/0210095.}

\bibitem{Klimcik:2008eq}
  C.~Klimcik,
{\it On integrability of the Yang-Baxter sigma-model},
  J. Math. Phys. {\bf 50} (2009) 043508,
  \href{http://arxiv.org/abs/0802.3518}{arXiv:0802.3518 [hep-th]}.

\bibitem{Arutyunov:2013ega}
  G.~Arutyunov, R.~Borsato and S.~Frolov,
  {\it S-matrix for strings on $\eta$-deformed $AdS_{5} \times S^5$,}
  JHEP {\bf 1404} (2014) 002,
\href{http://arxiv.org/abs/1312.3542}{arXiv:1312.3542 [hep-th].}
  
 



\end{thebibliography}
\end{document}